\documentclass[final,3p,onecolumn,authoryear,hidelinks]{elsarticle}
\usepackage{graphicx}
\usepackage{epstopdf}
\usepackage{amssymb}
\usepackage{amsmath}
\usepackage[unicode,pdfencoding=auto,psdextra,hidelinks,pdfhighlight=/P]{hyperref}
\usepackage{xspace}

\newcommand{\GG}[2]{\ensuremath{#1^\circ\,\text{#2}}}
\newcommand{\E}[1]{\GG{#1}{E}}
\newcommand{\W}[1]{\GG{#1}{W}}
\newcommand{\N}[1]{\GG{#1}{N}}

\newcommand{\NN}[2]{\GG{#1}{N}\,\text{--}\,\GG{#2}{N}}
\newcommand{\WW}[2]{\GG{#1}{W}\,\text{--}\,\GG{#2}{W}}
\newcommand{\EW}[2]{\GG{#1}{E}\,\text{--}\,\GG{#2}{W}}
\newcommand{\mdash}{\,--\,}

\newcommand{\chla}{chlorophyll~{\em a}\xspace}
\newcommand{\Chla}{Chlorophyll~{\em a}\xspace}

\hypersetup{linkcolor=Plum,citecolor=OliveGreen,urlcolor=NavyBlue,colorlinks=true}

\begin{document}
\begin{frontmatter}
\title{Mesoscale circulation in the Alaskan Stream area}

\author{S.V. Prants}
\ead{prants@poi.dvo.ru}
\ead[url]{http://dynalab.poi.dvo.ru}
\author{A.G. Andreev}
\author{M.Yu. Uleysky}
\author{M.V. Budyansky}

\address{Pacific Oceanological Institute of the Russian Academy of Sciences,\\
43 Baltiiskaya st., 690041 Vladivostok, Russia}

\begin{abstract}
The Alaskan Stream is the northern boundary current in the subarctic North Pacific.
This area is characterized by significant temperature, salinity and density
differences between  coastal and open-ocean waters and strong mesoscale dynamics.
In this paper we demonstrate the transport pathways of Alaskan Stream water in the eastern subarctic Pacific and
the eastern Bering Sea from October~1, 1994 to September~12, 2016 with the help of altimetry-based
Lagrangian maps. A mesoscale eddy activity along the shelf-deep basin boundaries in the Alaskan
Stream region and the eastern Bering Sea is shown to be related with the wind stress curl in the
northern North Pacific in winter. A significant correlation is found between the concentration of
\chla in the Alaskan Stream area and eastern Bering Sea in August\mdash September and the wind
stress curl in the northern North Pacific in November\mdash March. The mesoscale dynamics, forced
by the wind stress curl in winter, may determine not only lower-trophic-level organism biomass but
also salmon abundance/catch in the study area.
\end{abstract}

\begin{keyword}
Alaskan Stream \sep eastern Bering Sea \sep transport pathways \sep Lagrangian maps \sep mesoscale eddies \sep salmon catch
\end{keyword}
\end{frontmatter}

\section{Introduction}

\begin{figure}[!htp]
\begin{center}
\includegraphics[width=0.7\textwidth,clip]{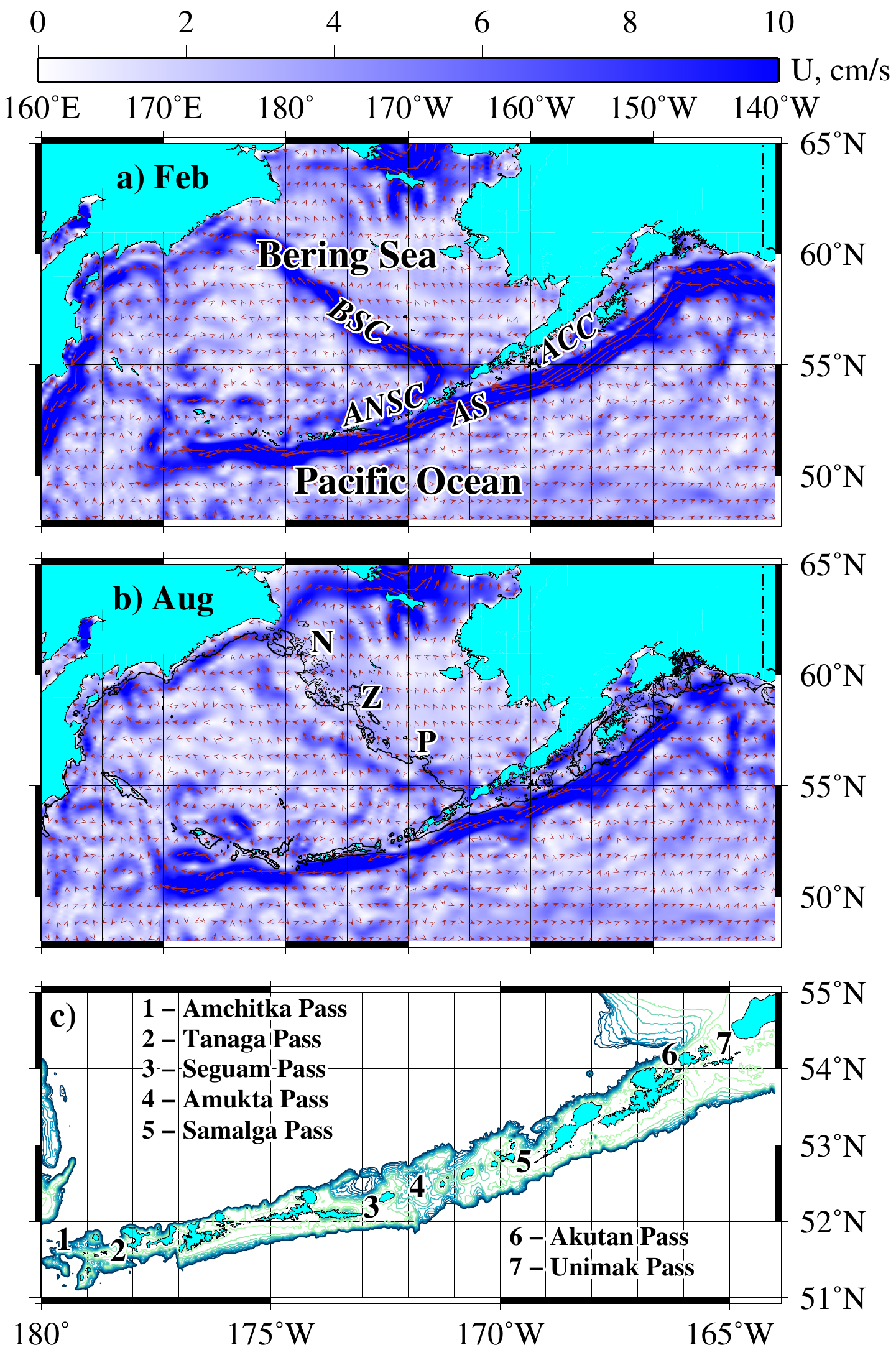}
\end{center}
\caption{The altimetric AVISO velocity field averaged for a) February and b) August from 1994 to 2016.
c) The Aleutian passes. Abbreviations: ACC~--- Alaska Coastal Current, ANSC~--- Aleutian North Slope
Current, AS~--- Alaskan Stream, BSC~--- Bering Slope Current, N~--- Navarin Canyon, Z~--- Zhemchug
Canyon and P~--- Pribilof Canyon.}
\label{fig1}
\end{figure}

The Alaskan Stream (AS) is the northern boundary current of the North Pacific Subarctic Gyre flowing
westward along the shelf-break to the south of the Alaska Peninsula and the Aleutian archipelago.
The AS is a narrow (${<}100$~km), deep (${>}5000$~m) and high-speed current with the transport of the
order of $14\text{--}40 \cdot 10^6$~m$^{3}$~s$^{-1}$ \citep{Favorite1974,Ueno2010}. Portions of the AS
flow through the Aleutian passes form the Aleutian North Slope Current (Fig.~\ref{fig1}), the narrow
and high speed current that flows northeastward along the north slope of the Aleutian Islands
\citep{Stabeno2009}, and the Bering Slope Current flowing north-westward along the eastern shelf-break
of the Bering Sea (BS) \citep{Favorite1974, Stabeno_Reed_1994, Johnson2004}. The northward flow of the
AS water through the Aleutian passes is a main source of nutrients and heat for the BS ecosystem
\citep{Stabeno05}. The variations in the Alaska Gyre waters supply, caused by the AS, lead to
interannual variations in the dissolved oxygen and temperature in the intermediate layer of the
Okhotsk Sea and western subarctic Pacific area \citep{Andreev_2006}. Enhancement of the AS flow
is accompanied by an increase in sea surface temperature and decreasing ice area in the Okhotsk
Sea in winter and can be considered as direct and indirect causes of a reduction in the \chla
concentration and large-sized zooplankton biomass in the eastern Okhotsk Sea in
winter-spring \citep{Prants2015a}.

Mesoscale variability is an important factor in the eastern subarctic Pacific and BS dynamics
\citep[e.g.,][]{Okkonen_2001, Okkonen2003, Okkonen2004, Ladd2007}. Mesoscale eddies enhance
the cross-shelf exchange of macronutrients, iron and phytoplankton and zooplankton populations
\citep[e.g.,][]{Johnson2005, Okkonen2003}. \citet{Crawford2005, Crawford2007, Ueno2010, Brown2012}
have indicated that such eddies play a significant role in controlling time and space patterns of
\chla  and may, therefore, determine the biological productivity and ecological function in the
region. Interannual and decadal modulations of the eastern subarctic Pacific open-ocean ecosystems
may be explained by analyzing statistics of eddy-induced cross-shelf transport \citep{Combes2009}.
The impact of mesoscale eddies on the circulation and biology in the eastern BS have been examined
by many authors \citep[e.g.,][]{Okkonen2004, Mizobata02, Mizobata2006, Ladd_Stabeno_Ohern_2012}.
Instabilities in the Bering Slope Current, wind forcing, topographic interactions and flow through
the eastern Aleutian passes have been suggested to be possible eddy-generation mechanisms.
Analysis of the SSH time-series in 2002--2012 at the eastern boundary of the subarctic
gyre demonstrated that the year-to-year changes of the SSH in the anticyclonic eddies were
related to the wind stress curl in winter. It was assumed that spin up of the subarctic cyclonic
gyre, forced by the wind stress curl, may enhance the anticyclonic eddy activity in the AS
area \citep{Prants2013}.

In this study we focus at the AS transport pathways by using altimetry-based Lagrangian maps
and forcing patterns that contribute to the interannual variability of the mesoscale dynamics
and the \chla concentration in the east BS and AS area. We show that the intensity of AS
anticyclonic eddies and anticyclonic eddies in the south-eastern BS is determined by the
wind stress curl (WSC) in the northern North Pacific in November\mdash March. There is a
significant correlation between the concentration of \chla at the deep basin margins in the
BS and eastern subarctic Pacific in August\mdash September and the WSC in the northern North
Pacific in winter. Our results indicate that mesoscale dynamics in the eastern BS and AS areas
may determine not only lower-trophic-level organism (the autotrophic phytoplankton) biomass
but also the salmon abundance/catch.

The paper is organized as follows. Section~2 describes briefly the data we use and the Lagrangian
methods we apply to study transport pathways, origin, history and fate of different water masses.
The next section~3 with the main results consists of two parts. Firstly, we study a correlation
between the WSC in the northern North Pacific in winter and  mesoscale eddy activity along the
deep basin boundaries in the AS region and the eastern BS. The altimetry-based Lagrangian simulation
is used to track the penetration of AS and open-ocean waters into the eastern BS. Secondly, we study
impacts of the mesoscale activity on \chla concentration in the area and its correlation with
salmon abundance and catch. Section~4 discusses possible physical mechanisms
of the correlations found.

\section{Data and methods}

Geostrophic velocities and sea surface heights (SSHs) were obtained from the AVISO database
(\url{http://www.aviso.altimetry.fr}) archived daily on a $1/4^{\circ}\times 1/4^{\circ}$ grid from
October~1, 1994 to September~12, 2016. The distributed global product combines altimetric data from
the TOPEX/POSEIDON mission, from Jason-1 for data after December 2001 and from Envisat for data
after March 2002. The meridional and zonal velocities and SSHs are gridded on a
$1/4^{\circ}\times 1/4^{\circ}$ Mercator grid, with one data file every day.

To compute the WSC
($\operatorname{curl}_{z}({\partial \tau_{y}}/{\partial x}-{\partial \tau_{x}}/{\partial y})$,
where $\tau_{y}$ and $\tau_{x}$ are respectively meridional and zonal wind stress components in
the northern North Pacific) we used the monthly wind stress dataset from the NCEP reanalysis.
Ocean \chla concentration data have been obtained from monthly composites from MODIS satellite
imagery of ocean color (Level-3 product) with a horizontal resolution of 9~km on a regular grid
(\url{http://oceancolor.gsfc.nasa.gov}). The salmon catch statistic was downloaded from the North
Pacific Anadromous Fish Commission website (\url{http://www.npafc.org}). ARGO floats data (tracks,
seawater temperature and salinity) and bottle oceanographic data (temperature, salinity, nutrients
and \chla concentration) have been provided by the National Oceanographic Data Center
(\url{http://www.nodc.noaa.gov}).

Lagrangian maps are geographic plots of Lagrangian indicators versus simulated particle's initial
positions. We used before as Lagrangian indicators different functions of particle's trajectory
\citep{FAO13,Springerbook2017}. They have been shown in recent years to be useful for studying
large-scale transport and mixing in the ocean including propagation of radionuclides in the western
North Pacific after the accident at the Fukushima Nuclear Power Plant \citep{DSR2015} and finding
of potential fishing grounds \citep{Prants2014c}. In order to track penetration of the AS and
open-ocean waters into the eastern BS, we use in this paper a special kind of the Lagrangian maps
which we call the origin maps. Being computed backward in time, the origin maps show where the
waters came from to a study area and from which water masses a studied eddy consists of.
They are computed as follows.

The vast area in the northern North Pacific, \NN{50.0}{65.0}, \EW{160.0}{145.0}, is seeded each
three days with a large number of virtual particles for the period of time from October~1,
1994 to September~12, 2016. Their trajectories in the altimetric AVISO velocity field are
computed backward in time for a year solving advection equations for passive particles with
a fourth-order Runge\mdash Kutta scheme
\begin{equation}
\frac{d \lambda}{d t} = u(\lambda,\varphi,t),\qquad \frac{d \varphi}{d t} =v(\lambda,\varphi,t),
\label{adveq}
\end{equation}
where $u$ and $v$ are angular zonal and meridional altimetric geostrophic velocities, $\varphi$
and $\lambda$ are latitude and longitude, respectively. Bicubical spatial interpolation and third
order Lagrangian polynomials in time are used to interpolate the velocity field.

We are interested in three water masses and their transport pathways. To track the AS waters, the
section along the meridian $x_0=\W{145}$ from $y_0=\N{58}$ to $y_0=\N{60}$ is fixed. The particles,
which crossed that section in the past, are colored in red on the origin Lagrangian maps. The open-ocean
particles, which crossed the section $x_0=\EW{160.0}{164.0}$, $y_0=\N{50.0}$ in the past, are colored
in green. The eastern BS particles, which crossed the section from \E{177.0}, \N{62.0} to \W{164},
\N{55.0} in the past, are colored in blue (see the yellow line Fig.~\ref{fig2}a). We removed from
consideration all the particles entered into any AVISO grid cell with two or more corners touching
the land in order to avoid artifacts due to the inaccuracy of the altimetry-based velocity field
near the coast. Due to impact of the coast, the Alaskan Coastal Current pathways in the Pacific
Ocean and BS \citep[see, e.g.,][]{Schumacher1989,Stabeno05} have not been analyzed using
altimetry-based Lagrangian maps. The corresponding colored Lagrangian maps demonstrate
clearly transport pathways of those water masses in the study area.

It is useful to identify locations in the ocean with zero geostrophic velocity. In the theory
of dynamical systems, these can be termed ``elliptic and hyperbolic stagnation points'' that
are indicated on the Lagrangian maps by triangles and crosses, respectively. The elliptic points,
located mainly in the centers of eddies, are those points around which the motion of water is
stable and circular. The upward oriented blue triangles mark centers of anticyclones and downward
oriented triangles in red indicate cyclones. The hyperbolic points, located mainly between and
around eddies, are unstable stagnation points with the direction along which water parcels converge
to such a point and another direction along which they diverge. The existence of hyperbolic areas in
the ocean have been recently confirmed by tracks of drifters
in the western North Pacific \citep{Prants2016}.

\section{Results}
\subsection{The Alaskan Stream eddies and transport pathways of the Alaskan Stream
and open-ocean Pacific waters into the eastern Bering Sea revealed by the Lagrangian maps}

\begin{figure*}[!htp]
\begin{center}
\includegraphics[width=0.9\textwidth,clip]{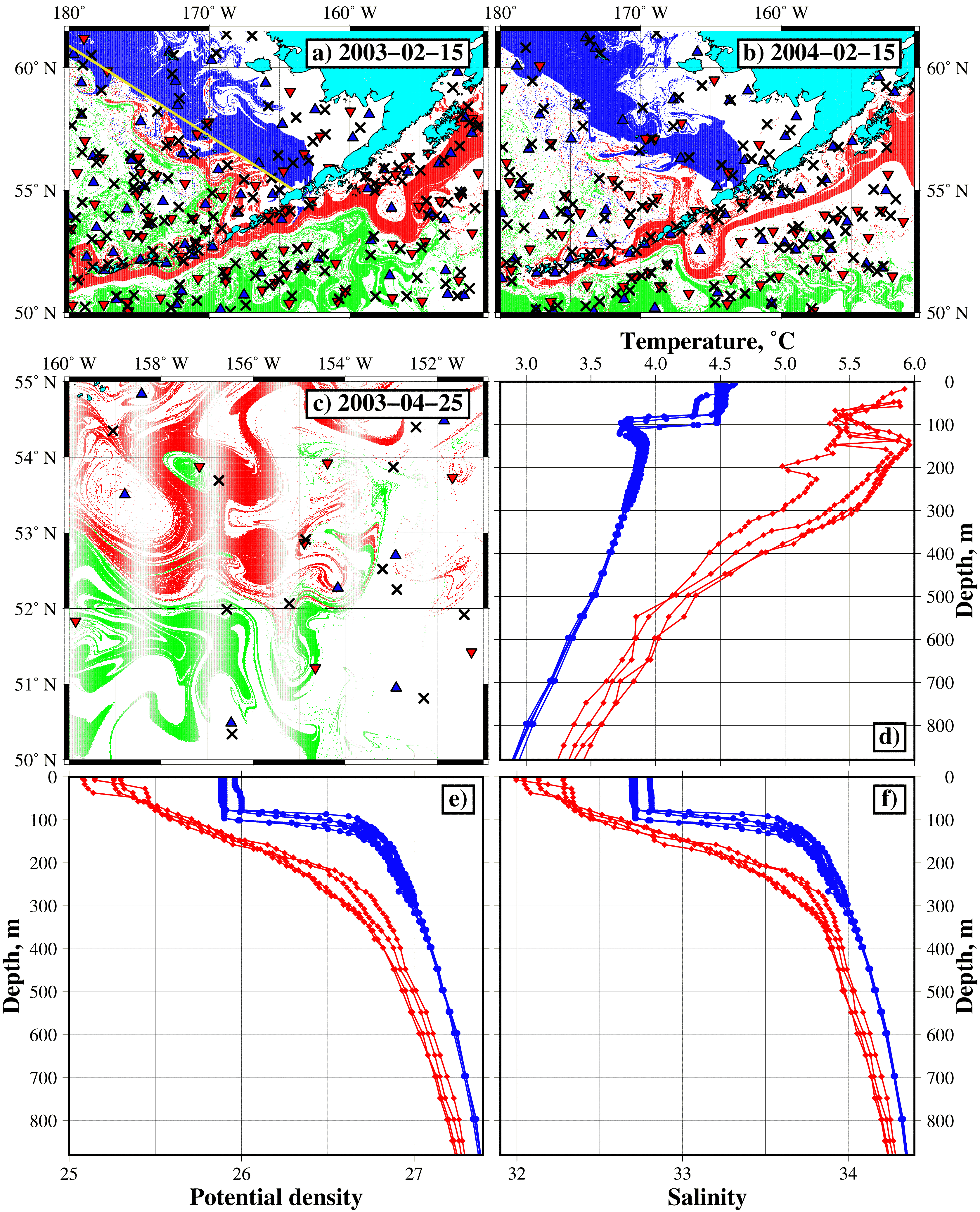}
\end{center}
\caption{a) and b) The origin Lagrangian maps show the intrusion of AS waters (red) and off-shore
subarctic waters (green) from the northern Pacific into the eastern BS through the Aleutian passes
in February 2003 and February 2004. The penetration of the BS shelf waters into the deep basin of
the eastern BS is demonstrated by the blue color. c) The map on April 24, 2003 with the AS anticyclone
ASAC~2003--2004 centered at \N{53.5}, \W{158.8} and the eastern subarctic cyclone (the green patch
of open-ocean water centered \N{51.2}, \W{154.6}). d)--f) The vertical distributions of temperature,
salinity and relative density in the ASAC~2003--2004 (the red profiles) and the subarctic cyclone with
open-ocean water (the blue profiles) in March\mdash May 2003 (data from the ARGO buoys nos.~49070 and
4900176). Elliptic (stable) stagnation points with zero mean velocity at the fixed date are indicated
by the downward and upward oriented triangles which mark cyclones and anticyclones, respectively.}
\label{fig2}
\end{figure*}

\begin{figure*}[!htp]
\begin{center}
\includegraphics[width=0.93\textwidth,clip]{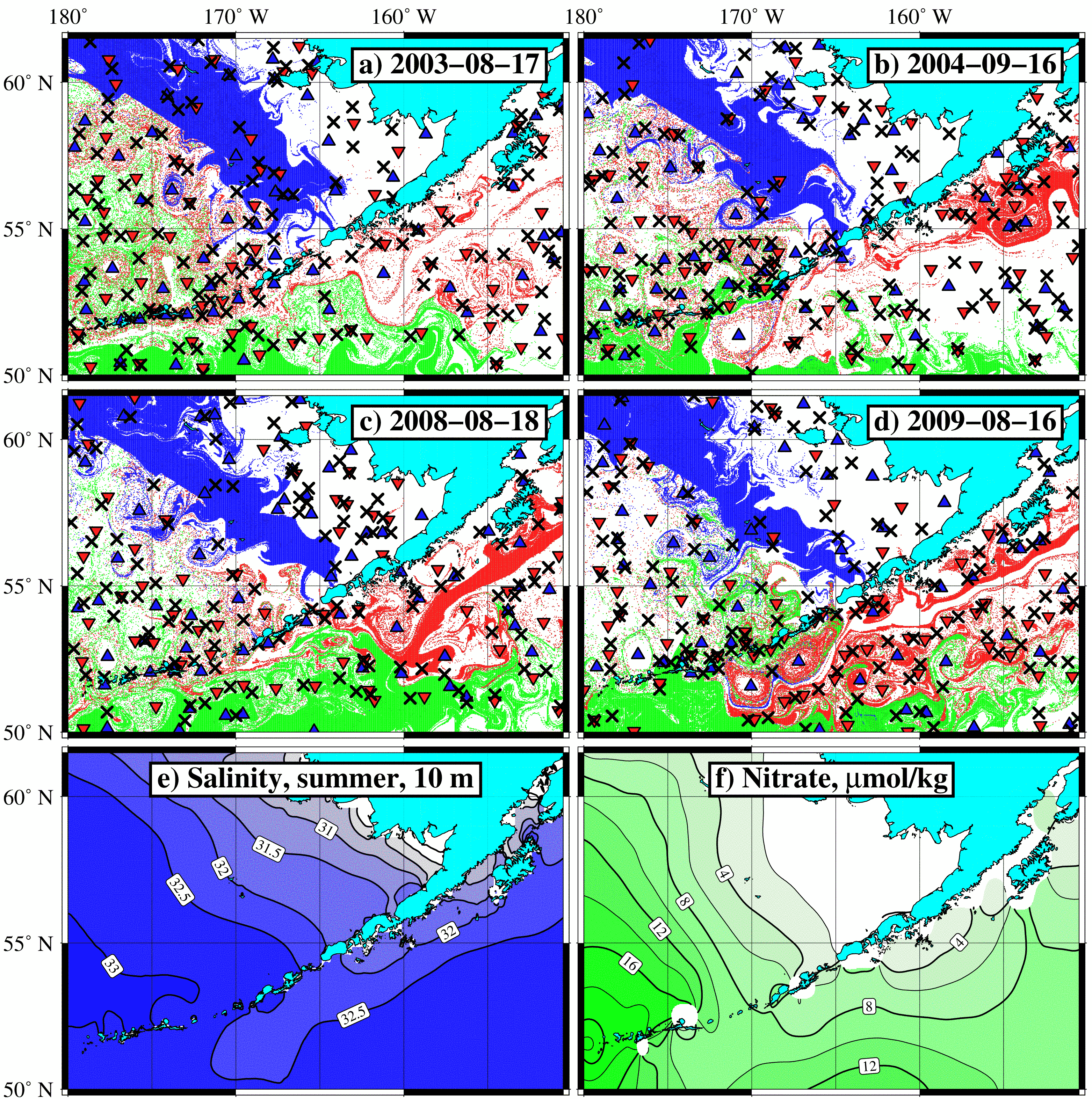}
\end{center}
\caption{a)--d) The origin Lagrangian maps demonstrate the impact of the AS anticyclonic eddies on
the supply of ``red'' AS waters and ``green'' open-ocean subarctic waters from the northern Pacific
into the eastern BS and penetration of the ``blue'' BS shelf waters into the deep BS basin in August
and September 2003, 2004, 2008 and 2009. e) and f) The distributions of the salinity and nitrate
concentrations in the surface waters (10~m) of the northern Alaska and eastern BS in summer
(July\mdash September, World Ocean Atlas 2013, $1^\circ$ grid).}
\label{fig3}
\end{figure*}

The strong and narrow AS in the northern Pacific, the Aleutian North Slope Current and the Bering
Slope Current in the BS are clearly visible in the altimetric AVISO velocity field averaged for February
(Fig.~\ref{fig1}a) from 1994 to 2016. The main inflow of the AS waters into the BS occurs through the
Amchitka (\W{180}) and Amukta (\W{172}) Passes. In August, the surface circulation in the eastern BS
is determined by  mesoscale anticyclonic and cyclonic activity. The mesoscale anticyclones in the area
of the Bering Slope Current are topographically constrained with the Navarin, Zhemchug and Pribilof
canyons (Fig.~\ref{fig1}b). An intensification of the southwestward flow of AS water in the northern
North Pacific and the north-northwestward flow of AS water in the BS are typically observed in
November\mdash March when the Aleutian Low pressure cell is activated. Off-shore AS meanders are
related to the mesoscale anticyclonic and cyclonic eddies. The origin Lagrangian maps in
Figs.~\ref{fig2}a--c and \ref{fig3}a--d and the surface salinity and nitrate distributions in
Figs.~\ref{fig3}e and~f show that less saline and relatively low nitrate AS waters (marked by
the red color) intrude into the BS through the Aleutian passes and then flow northwestward along
the Bering slope. The small and large eddies in the Aleutian Islands area stimulate inflow of
the AS water and the open-ocean subarctic water (marked by green) into the BS.

The southwestward drift of the AS mesoscale anticyclones along the Alaskan Peninsula and eastern
Aleutian Islands forced the AS water southward into the deep Pacific basin and enhanced the AS and
open-ocean water to inflow in the BS (Figs.~\ref{fig1}a and~b, Fig.~\ref{fig3}a--d). We focused on
the mesoscale anticyclonic eddies originating in the northern part of the Gulf of Alaska and advected
by the AS along the Pacific Ocean side of the eastern Aleutian Islands. These eddies have typically
an elliptical shape with a dimension of about $150\times 200$~km with the centers (the elliptic points)
located over the axis of the Aleutian bottom trench. Inspecting the daily Lagrangian maps, we have
found that one of such eddies, which we call ASAC~2003--2004, originated in the northern part of
the Gulf of Alaska in winter 2002.
% Элиптическая точка вихря образовалась в точке \N{59}, \W{146} 2002-01-22).
In summer and fall 2002, it was relatively weak and located to the southwest off the Kodiak
Island with the  center at around \N{55.5}, \W{153}. The reinforcement and enlargement of this
eddy occurred in January\mdash March 2003 southward of the Alaskan Peninsula in the area \NN{53}{55},
\WW{156}{158} when the WSC over the northern North Pacific increased significantly.

The ASAC~2003--2004 is clearly seen on the origin Lagrangian map in Fig.~\ref{fig2}a with the center
at around \N{54.2}, \W{157} on February~15, 2003. It consists of the core with a ``white'' water, a
periphery with the AS ``red'' water  surrounded by the ``green'' open-ocean waters. The intensification
of this eddy was accompanied by advection of the open-ocean water by a cyclonic eddy toward the western
and southern edges of the ASAC~2003--2004 (Figs.~\ref{fig2}a and~c) and thereby an increase of the
temperature, salinity and density differences between the ASAC~2003--2004 and outside open-ocean
waters (Figs.~\ref{fig2}d--f). The ARGO CTD data (buoys nos.~49070 and 4900176) demonstrate that the
ASAC~2003--2004 had a warmer, less-saline and less-dense core than the open-ocean subarctic water.
During its southwestward drift in 2004, the ASAC~2003--2004 controlled supply of the AS water
(Fig.~\ref{fig2}b) and open-ocean subarctic water into the BS through the Aleutian passes
(Fig.~\ref{fig3}b).

The eddy, which we call ASAC~2005--2006, was not advected to the \NN{53}{55}, \WW{156}{158} area in
winter 2004 during the period of low WSC in the northern North Pacific but stayed southward of Kodiak
Island and was relatively weak \citep{Ladd2007}. We observed its activation and reinforcement in winter
2005 when the WSC over the northern North Pacific increased. During 2005 and 2006, the ASAC~2005--2006
forced the transport of AS and open-ocean waters into the BS (see Figs.~\ref{fig1S}a and~b).
Figures~\ref{fig2S}a and~b in Appendix show the vertical distributions of temperature,
salinity and potential density in the ASAC~2005--2006 and outside waters in September 2005 and September
2006. Similar to the ASAC~2003--2004 (Figs.~\ref{fig2}d--f), the ASAC~2005--2006 core was composed of
relatively low salinity ($33.7\text{--}33.9$) and low density ($26.7\text{--}26.9$) waters.
The temperature of waters inside of the anticyclone was 1--2~$^\circ$C higher than outside it.

%

%
% ASAC~2008-2009 элиптическая точка этого вихря образовалась \W{149}, \N{58} 2007-01-04,
% Смещается на юго-задад где стациоирует в районе \W156-155 \N54-55 до 2008-02-15
% временами от него отщепляются АЦ, которые смещаются на запад не далее \W161.
% 2009-01-24 в точке \W{162}, \N{54} от вихря ASAC~2008-2009 отщепляется ``западный вихрь",
% который образует с ASAC~2008-2009 долгоживущую вихревую пару
%
% не понятно что имеется ввиду под AS water (summer 2008) and open-ocean pacific water (summer 2009)
%
An increase of the WSC in winter 2008 led to enlargement and strengthening of the eddy ASAC~2008--2009
in the \NN{53.5}{55}, \WW{156}{158} region and formation of a second anticyclonic eddy. During its
southwestward drift in 2008 and 2009, the anticyclonic eddies centered at \N{51.5}, \W{170}
and \N{52.4}, \W{167} in August~16, 2009 forced the intrusions of AS water (summer 2008) and
open-ocean Pacific water (summer 2009) into the BS (Figs.~\ref{fig3}c and~d). In winter 2010,
the eastern anticyclonic eddy was advected to the south of the Aleutian Islands and the western
eddy drifted along the Aleutian Islands to the western subartic Pacific. In fall 2010 and winter
2011, the ASAC~2008--2009 was observed in the central part of the Western Subarctic Gyre with
the elliptic point at \N{51}, \E{170}.
%
% 2010-07-22 западный вихрь и ASAC~2008-2009 расщепляются в районе \W173 \N51
% западный вихрь, затем смещается на юг и выходит из области наблюдения
% ASAC~2008-2009 постепенно смещается на запад, где 2010-07-22, \E{174}, \N{51},
% сливается (наматывается) на АЦ (который имеет местное происхождение,
% образовался ранее 2010-01-31 в точке \E{172} \N{52})

The supply of the AS water through the Aleutian passes leads to formation and strengthening of
anticyclonic eddies in the eastern BS. In January\mdash February 2003, an inflow of the AS water
through the Amchitka Pass led to formation of the anticyclonic eddy (the red patch centered at
around \N{52.5}, \W{179}) in the southern BS (Fig.~\ref{fig2}a). The generation of the ``Pribilof
mesoscale anticyclonic eddy 2004'' in the BS was observed in the eastern Aleutian Passes in spring
2004. In May 2004 it intensified probably due to a density difference at its edges between the ``red''
AS water and ``green'' open-ocean water. In June 2004 its center was at the point \N{54.5}, \W{168}
(see Fig.~\ref{fig1S}c). In July\mdash September it occupied its position to the
southwest of Pribilof canyon with the elliptic points at \N{55.5}, \W{171} (Fig.~\ref{fig3}b).
In October 2004 it moved westward to the deep BS. During its stay in the Bering Slope Current
region, the Pribilof mesoscale anticyclonic eddy 2004 trapped the ``blue'' BS shelf water and
wound around it (Fig.~\ref{fig3}b). The penetration of the BS outer shelf water into the eddy
at that place has been shown by the CTD observations conducted in June 1997
\citep{Ladd_Stabeno_Ohern_2012} (see Fig.~\ref{fig1S}d).
%
% Элиптическая точка ивхря образовалась 2004-05-09 в точке \W{168} \N{54}, после чего
% вихрь постепенно увеличиваясь в размерах смещается на северо-запад, в конце августа
% вихрь захватывает часть синих вод с шельфа, наматывая их на периферию
%
%
% координаты на 2004-09-16, вихрь постепенно смещается на запад перенося синюю воду не периферии
% его элиптическая точка аннигилирует 2004-12-11 в районе \W172, \N56, воды периферии наматываются
% на сосведние АЦ
%
\subsection{A correlation between the mesoscale eddy activity in the Alaskan Stream region
and the eastern Bering Sea and the wind stress curl in the northern North Pacific in winter}

The strength and position of the Aleutian Low pressure cell are the main factors which determine
the circulation in the northern North Pacific. The strong Aleutian Low and positive WSC pattern
over the northern North Pacific in November\mdash March spin-up the subarctic cyclonic gyre
\citep[e.g.,][]{Ishi_2005}. One may assume that that spin-up results in more eddy activity at
its northern boundary (the AS region). The anticyclonic eddies contain a core of low-density
water that produces an upward doming of the sea surface detectable by satellite altimeters.

\begin{figure}[!htp]
\begin{center}
\includegraphics[width=0.8\textwidth,clip]{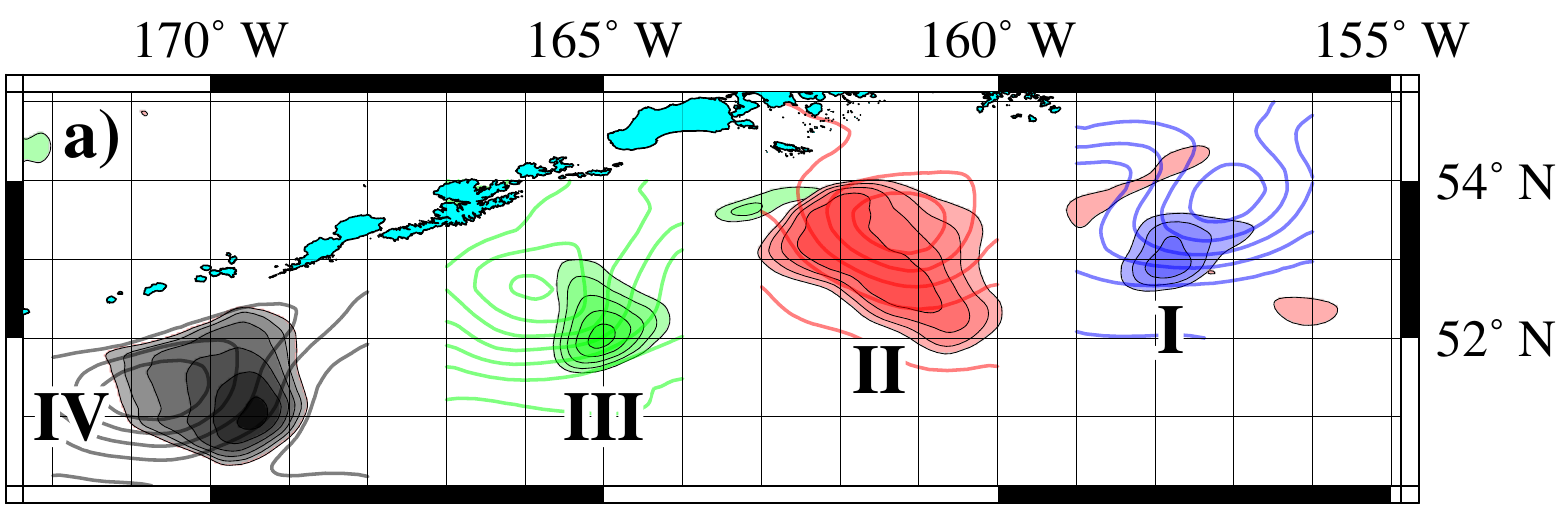}
\end{center}
\caption{The distribution of the correlation coefficients (0.60--0.90, int. = 0.05)
between the change of the WSC (November\mdash March) and SSH in the AS area in March
(I), August (II), February (III, 1-yr lagged WSC) and September (IV, 1-yr lagged WSC).
The SSH isolines in March and August, 2003 and February and September, 2004.}
\label{fig4}
\end{figure}

\begin{figure*}[!htp]
\begin{center}
\includegraphics[width=0.93\textwidth,clip]{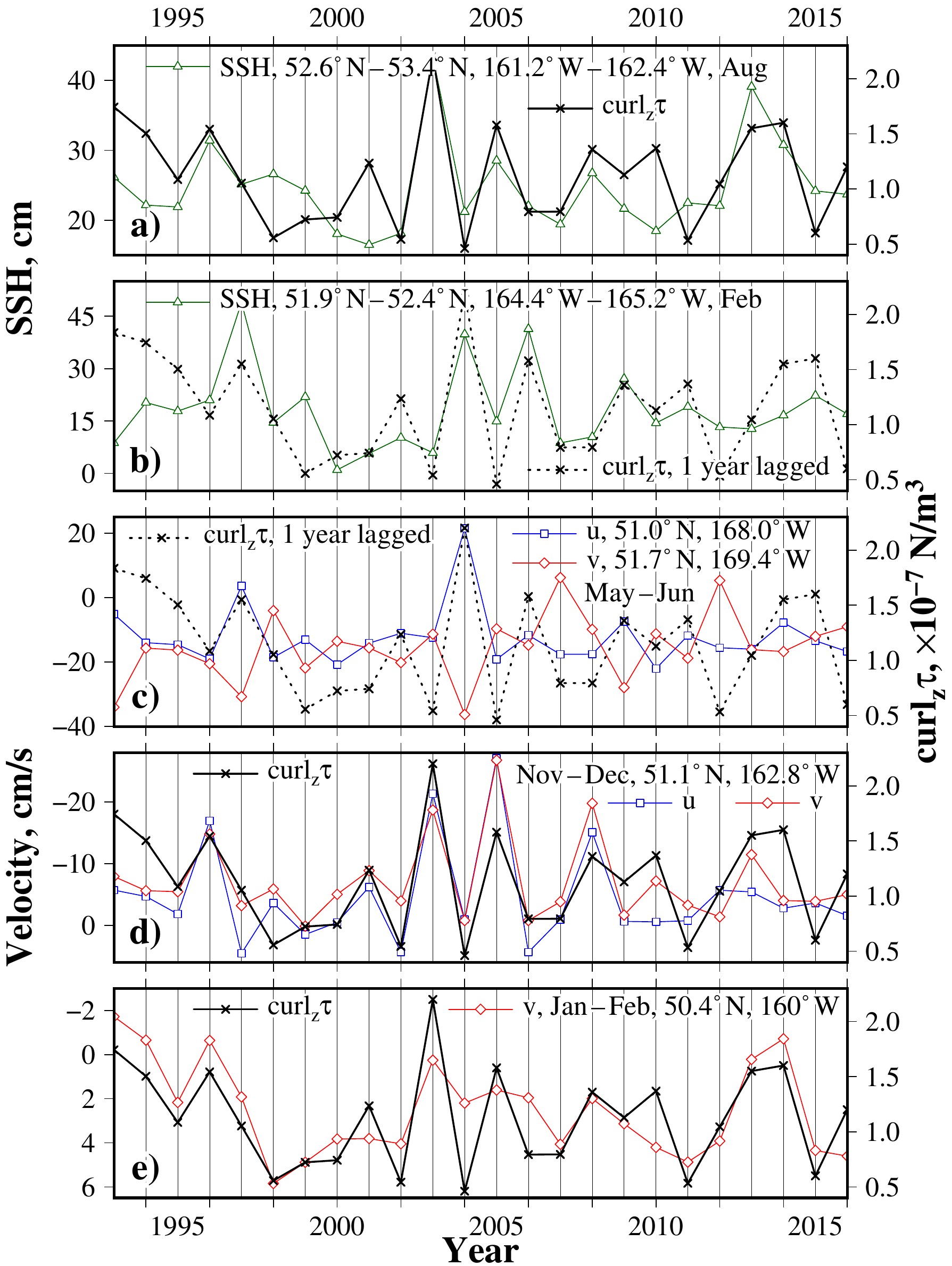}
\end{center}
\caption{a)--e) The year-to-year changes of the WSC (November\mdash March) in the
northern North Pacific, SSH, the meridional and zonal velocities in the AS area.}
\label{fig4be}
\end{figure*}

The increased (decreased) SSH in AS anticyclonic eddies are associated with the increased
(decreased) WSC in the northern North Pacific (\NN{46}{48}, \EW{165}{170}) in winter with
the correlation coefficient $r=0.60\text{--}0.90$, 2002--2016 (Fig.~\ref{fig4}). Amplitude
of the steric height variability in the study area is about 2.4~cm (the thermosteric
height signal is about 2~cm and halosteric height signal is about 0.4~cm) \citep{Qiu2002},
which is several times less than the amplitude of the interannual variations of SSH in the
AS area in February and August (around 20~cm) (Figs.~\ref{fig4be}a and b).

To find a correlation between the WSC and SSH and the  velocities at the boundaries of AS
anticyclonic eddies, we used monthly averaged SSH and velocities. If the correlation was found
to be significant for a two month period (May and June, July and August, etc.) we used the SSH
and velocities averaged for May\mdash June, July\mdash August, etc. (see the corresponding
figure captions or legends). There is a good correlation with $r=0.70\text{--}0.80$ between
the monthly averaged SSH along the northern and northeastern boundaries of the Gulf of Alaska
and monthly averaged zonal wind stress (\NN{55}{60}, \WW{140}{150}) in November\mdash March
when the Aleutian Low developed in the northern North Pacific \citep{Qiu2002}.
The formation of anticyclonic eddies along the northern boundary of the Gulf of Alaska
can be related to the along-shore wind and downwelling \citep[see, e.g.,][]{Combes2007}.
However, our results show that reinforcement of the AS anticyclonic eddies southward of the
Alaskan Peninsula (\NN{53}{55}, \WW{156}{158}) occurs during the periods of increased WSC
in the northern North Pacific. The correlation between the SSH in the AS anticyclonic eddies
and the WSC is significant for two years (Fig.~\ref{fig4}).

Our results demonstrate that the meridional and zonal velocities at the AS eddy boundaries
during two years (while the eddies drift southwestward along the western Alaska Peninsula
and eastern Aleutian Islands) are determined by the WSC in winter (Figs.~\ref{fig4be}c and~d).
The increased WSC in the northern North Pacific enhances the northward flow of the open-ocean
subarctic water to the southern boundary of the AS anticyclonic eddies in January\mdash February
(Fig.~\ref{fig4be}e) and thereby increases the density gradient at the eddy edges.

The changes in the Aleutian Low activity and WSC in the northern North Pacific in winter
determine year-to-year changes in velocities in some areas of the eastern BS. An increase
(decrease) of the WSC in the North Pacific in November\mdash March is accompanied by increased
(decreased) velocities at the boundaries of the anticyclonic eddies in the central part of the
deep BS in summer and fall (Fig.~\ref{fig3S}a). An intensification of the Aleutian Low and a large positive
WSC result in increasing of the northward flow on the BS outer shelf in the areas located close to
the Pribilof, Zhemchug and Navarin canyons (Fig.~\ref{fig3S}c). An increase (decrease) of the WSC in the
northern North Pacific in November\mdash March with a 1-year lag is accompanied by increased
(decreased) velocities at the boundaries of the anticyclonic eddies located in the area of
Aleutian North Slope Current in summer and fall (Fig.~\ref{fig3S}b).

\subsection{\Chla concentration and fish abundance and catch in the area}

Mesoscale eddies modulate primary production along the eastern subarctic Pacific shelf break
entraining \chla- and iron-rich shelf water while simultaneously transporting nitrate- and
silicate-rich basin water to the shelf \citep[see, e.g.,][]{Okkonen2003, Ladd2005, Crawford2007}.
The dominant feature in \chla distribution in the surface layer in the AS region is a contrast
between coastal and offshore waters. The coastal waters are productive with high values
of \chla concentration (${>}3$~$\mu$g/l in August and September), whereas the offshore
waters are oligotrophic with low \chla concentration values (${<}1$~$\mu$g/l).

\begin{figure*}[!htp]
\begin{center}
\includegraphics[width=0.93\textwidth,clip]{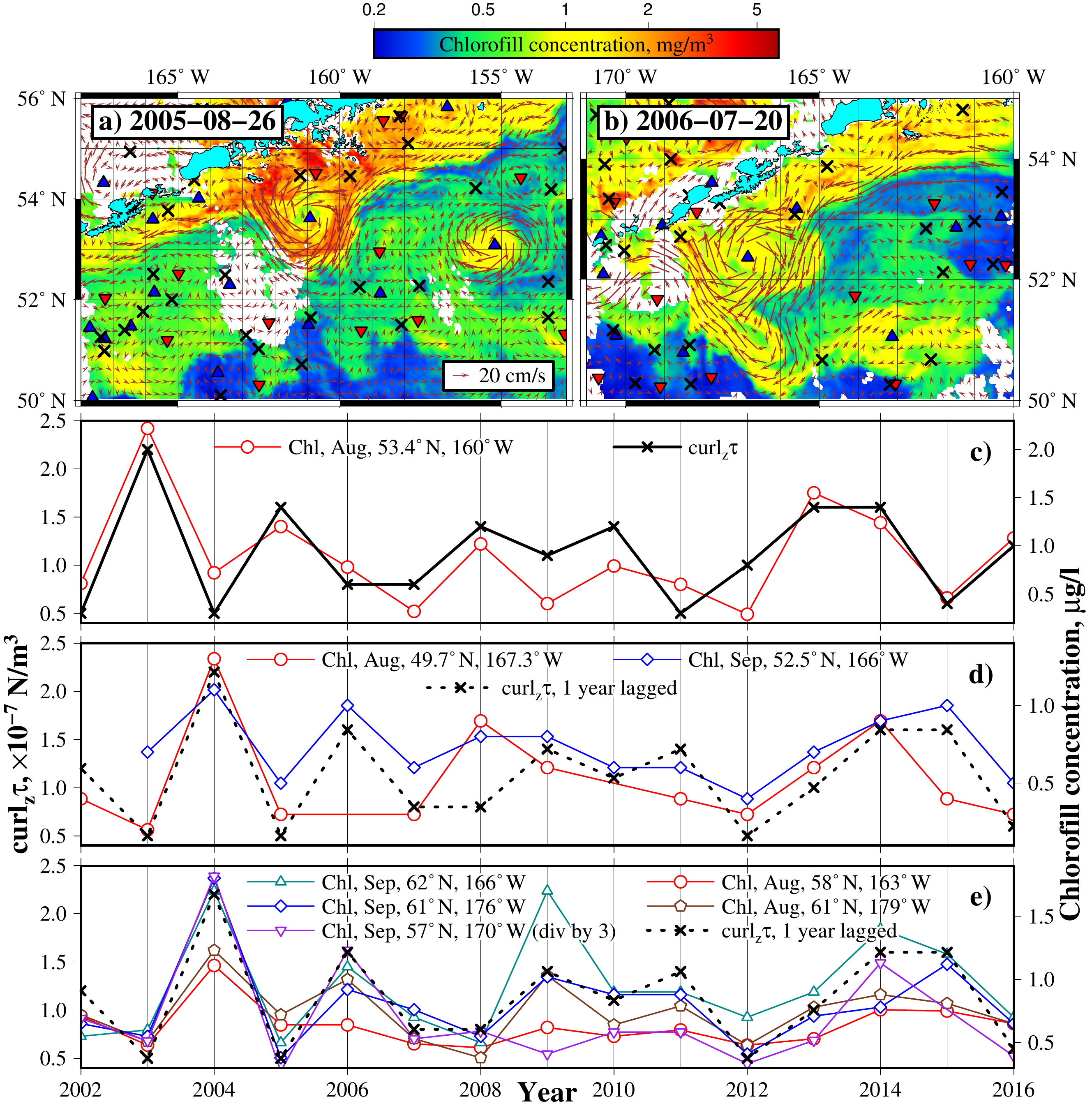}
\end{center}
\caption{a) and b) Distributions of the \chla concentration (the MODIS data) imposed on
the AVISO velocity field on the days indicated. c)--e) Year-to-year changes in the WSC
(November\mdash March) in the northern North Pacific and in the \chla concentration in the AS
and BS area. Elliptic and hyperbolic (unstable) stagnation points with zero mean velocity are
indicated by triangles and crosses, respectively.}
\label{fig6}
\end{figure*}

The \chla pool south of the Alaska Peninsula in summer 2005 and south of the eastern Aleutian
Islands in summer 2006 was affected by the mesoscale anticyclonic eddies centered at \N{53.5},
\W{161} and \N{52.5}, \W{167}, respectively (Figs.~\ref{fig6}a and~b). The filaments with high
\chla concentration ($1\text{--}2$~$\mu$g/l) have been transported off the shelf, wrapping around
the mesoscale eddies and than being trapped inside the eddies. The WSC forcing in winter modulates
the strength of anticyclonic eddies in the AS area and the velocities at their boundaries.
Year-to-year variations in location and strength of those anticyclonic eddies can determine
spatio-temporal changes in \chla concentrations in the surface waters in the AS region during
August\mdash September. Figures~\ref{fig6}c and~d show year-to-year changes of the \chla
concentration in August\mdash September and the WSC in the northern North Pacific in
November\mdash March. Large (in 2003 and 2005) and small (in 2002 and 2004) values of
the WSC in the northern North Pacific and, therefore, increased (decreased) velocities
at the eddy's boundaries (Figs.~\ref{fig4be}c and~d) have been accompanied by increased
(decreased) \chla concentrations at the boundaries of the AS eddies with no lag
(Fig.~\ref{fig6}c) and a 1-year lag (Fig.~\ref{fig6}d).

Figures~\ref{fig4S}a and~b show the distribution of \chla concentration
in the surface layer in the northern North Pacific and in the BS in August and September
2004 and a difference in the \chla concentration between 2004 and 2003. Large values of
the WSC over the northern North Pacific in winter 2003 resulted in strengthening of the AS
anticyclonic eddy and change in the velocity field in the eastern BS in 2003 and 2004
while that eddy was drifting along the Aleutian Islands. In August and September 2004,
the spots with high values ($1.5\text{--}3$~$\mu$g/l) of satellite-measured \chla have
been observed in the region of the Aleutian North Slope Current and along the shelf
break of the eastern BS (Figs.~\ref{fig4S}a and~b). In August and
September 2004, the surface waters have been significantly enriched by the \chla pigment
as compared to August and September 2003. In the central BS, we could see the shelf-break
front marking the boundary between low surface \chla and relatively fresh outer shelf
water and relatively high surface \chla and more saline basin water. This front is biologically
significant because it coincides with the BS ``Green Belt'', a region with high primary
production that supports an extensive variety of consumer species
\citep[see, e.g.,][]{Springer1996, Okkonen2004}).

Year-to-year changes in the \chla concentrations in the upper surface layer in the eastern BS
(similar to the AS region) have been positively correlated ($r=0.7\text{--}0.8$, 2002--2016)
with the WSC in the northern North Pacific in November\mdash March (Fig.~\ref{fig6}c).
An intensification of the Aleutian Low in winter and thereby an increase of the WSC in
the northern North Pacific were accompanied by increased \chla concentration in the
surface waters at the BS shelf edge with a 1.5-year lag).

Biomass of autotrophic plankton and concentration of \chla are determined by many factors,
such as solar radiation, seawater temperature, macro- and micro-nutrients availability,
water column stratification, etc. One of the most important factors, limiting phytoplankton
growth in the upper layer in the subarctic North Pacific and the BS in the post-spring-bloom
period (July\mdash September), can be a low supply of nutrients with dissolved inorganic nitrogen
and silicate considered to be the dominant elements limiting phytoplankton growth. In summer,
there is a good agreement between the spatial distributions of salinity and the nitrate
concentration in the surface layer in the BS and the eastern subarctic Pacific
(Figs.~\ref{fig3}e and~f). Low salinity (less than $32.2$) Alaskan waters are associated
with low nitrate concentrations (less than $4\text{--}6$~$\mu$mol~kg$^{-1}$) in the upper
surface layer. In the subarctic North Pacific and the BS, the vertical distribution of salinity
determines a density stratification in the upper layer. Existence of the strong vertical gradient
of salinity (a halocline) limits vertical exchange between the surface and deeper layers. Due to
tidal mixing in the Aleutian passes (such as the Seguam and Tanaga Passes) \citep{Stabeno05},
salinity and the concentration of nitrate in the upper surface waters in the BS deep basin reach,
respectively, $33$ and $15\text{--}20$~$\mu$mol~kg$^{-1}$. The nutrients, introduced due to mixing
in the passes and then advected northward, are critical to the BS ecosystem.

The vertical profiles of the \chla concentration in the eastern BS in summer are characterized
by a subsurface maximum located in the $10\text{--}40$~m layer (Fig.~\ref{fig4S}d). The concentrations
of \chla in the upper $40$~m layer reach $4\text{--}6$~$\mu$g/l. The depth of the \chla maximum
location is related to nutrient availability and a water column stratification. The observations,
conducted in July 2003 and 2004 in the eastern BS, demonstrate that the shallow ($10\text{--}20$~m)
\chla maxima have been related to the waters with strong salinity and density stratifications and
with salinities of $32.5\text{--}32.8$ in the surface layer. For relatively low ($32.3$) and high
surface salinities ($33.0$), the \chla maxima have been located in the $20\text{--}40$~m layer.
The shallow location of the \chla maximum provides higher \chla concentration in the
upper ($0\text{--}10$~m) surface layer captured by the satellite sensors.

Using the relationship between salinity and nitrate concentration \citep{Ladd2005,Mordy2005,Stabeno05},
the difference in satellite-measured \chla concentration in the eastern BS between summer 2003
and summer 2004 (Figs.~\ref{fig3S}a and~b) can be explained. In summer 2003,
the surface salinity was relatively low ($32.4$) in the most eastern part of the Bering
deep basin (\NN{54}{55}, \WW{170}{171}) and relatively high ($33.0$) in its southern part
(\NN{53}{54}, \WW{172}{174}) (Fig.~\ref{fig3S}e). In summer 2004, the distribution of salinity in
the surface layer in the eastern deep BS was quite uniform with the value of $32.8$ (Fig.~\ref{fig3S}e).

The salinity distributions in the deep eastern BS in summer 2003 and summer 2004 (Fig.~4Se)
are in a good agreement with the water mass distributions demonstrated by the origin Lagrangian
maps in Figs.~\ref{fig3}a and b. The Lagrangian maps show that in summer 2003 the southeast
of the BS was occupied by the ``green'' open-ocean subarctic (more saline) waters and
populated mainly by cyclones, while the BSC area was occupied by the ``red'' AS (less saline)
waters and populated by anticyclonic eddies (Fig.~\ref{fig3}a). In summer 2004, the deep
eastern BS was composed of spots of open-ocean and AS waters (Fig.~\ref{fig3}a) and was
characterized by a quite uniform salinity distribution possibly due to the anticyclones
eddies activity (Fig.~\ref{fig4S}e). The mixing, induced by anticyclonic eddies between the low
salinity coastal water and high salinity deep basin water, probably create favourable
conditions with nitrate availability and a shallow pycnocline for phytoplankton growth
and, thereby, significantly increased the \chla concentration in the upper surface layer
in the eastern BS in summer 2004.

The impact of the anticyclonic eddies on the \chla distribution in the AS area can be
demonstrated by using a Lagrangian indicator $L=\int\limits_{0}^{T} \sqrt{u^2+v^2}dt$, which
is a measure of a distance passed by advected  particles. The indicator $L$ is more suitable
for detecting and documenting vortex structures than the Lyapunov exponent and displacement
of particles $D$ from their initial positions \citep{Prants2016}. A studied area has been
seeded at a fixed date with a large number of virtual particles whose trajectories have been
computed backward in time for a month in the AVISO velocity field. The $L$ maps visualize not
only the very vortex structures but also a history of water masses to be involved in the vortex
motion in the past.

%The Lagrangian technique is based on the derivation of Lagrangian diagnostics from a
%time-dependent geostrophic velocity field.
%, and more precisely transport barriers and trajectories of synthetic passive tracers.
%A passive tracer initialized nearby the hyperbolic point is affected at the same time
%by contraction along the converging direction (called stable manifold of the hyperbolic point) and stretching along the diverging direction
%(unstable manifold). Lagrangian technique is well suited for diagnosing properties of tracers
%like chlorophyll, since they allow quantifying the dynamical properties experienced by a parcel of water during its motion.
%The orientation of the intrusion filaments (of either chlorophyll-rich or poor water) is in much better agreement
%with the unstable manifolds than with the instantaneous velocity field \citep[see, e.g.,][]{Lehahn07, Samuelsen2012}.

\begin{figure*}[!htp]
\begin{center}
\includegraphics[width=0.93\textwidth,clip]{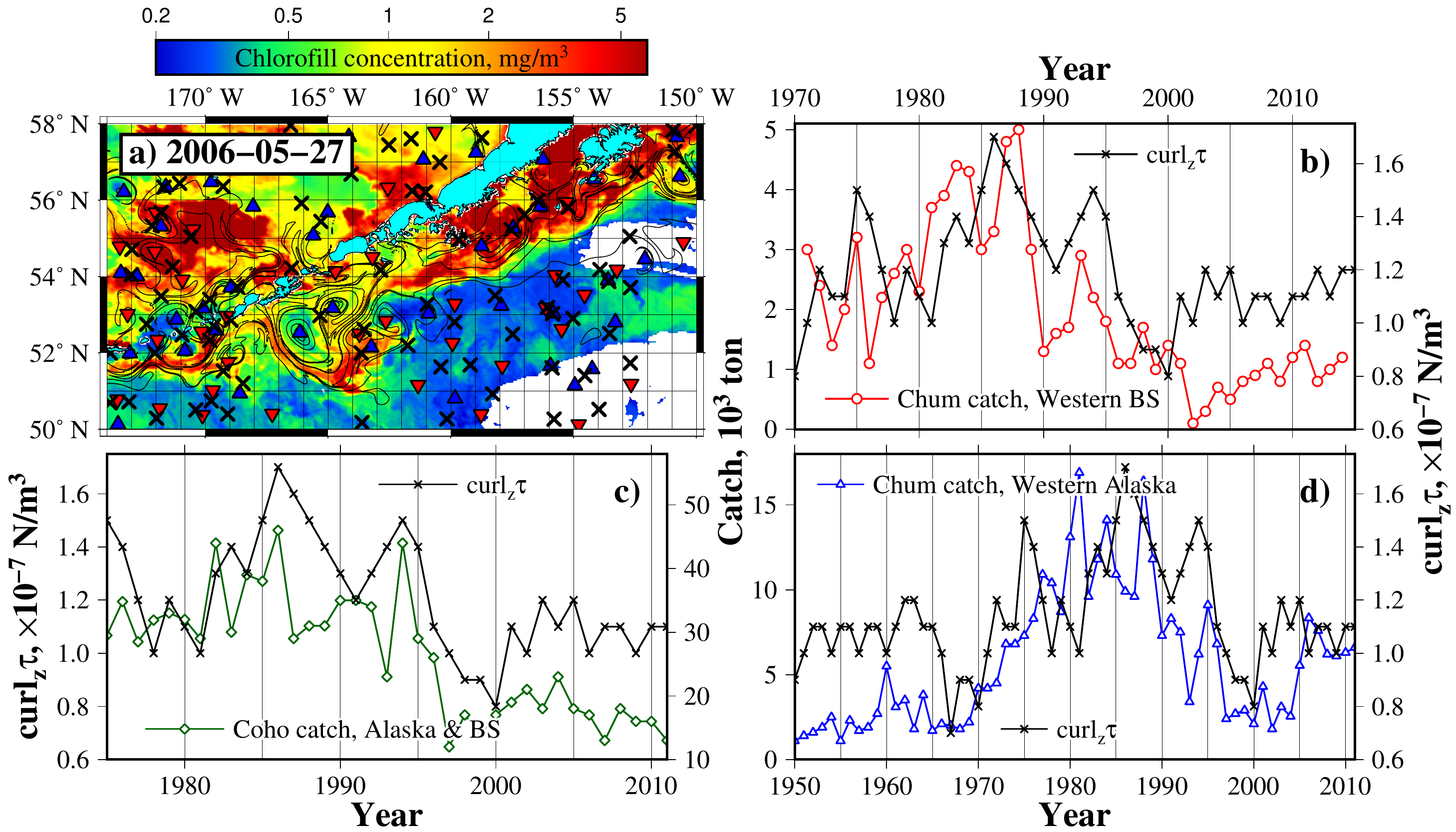}
\end{center}
\caption{a) The impact of anticyclonic eddies on the \chla distribution (the MODIS data)
in the AS area. The black contours are isolines of the Lagrangian indicator $L$ with the
step of 200 geographic minutes. Year-to-year changes in the WSC in November\mdash March
(the 5-years running mean) in the northern North Pacific and in the annual catch of b)
and d) chum salmon and c) coho salmon in the eastern subarctic Pacific and BS.}
\label{fig7}
\end{figure*}

In Figures~\ref{fig7}a and~\ref{fig5S} we impose distributions of the values of
the Lagrangian indicator $L$ in the AS area on the \chla patterns in May 2006, May 2010 and May
2011 with an intermittency of productive coastal waters with high \chla values (${>}6$~$\mu$g/l)
and oligotrophic offshore waters with low \chla  values (${<}1$~$\mu$g/l). The black contours in
those figures are isolines of $L$ with the step of 200 geographic minutes. They enclose stable
mesoscale eddies, such as ones centered at \N{52.5}, \W{165} and at \N{53}, \W{164}.
Filaments with high \chla concentration are transported offshore and wrapped
around persistent mesoscale eddies existing in the area.
%
% Весь текст этого параграф основан на рис.8, который призван показать
% выстраивание филаментов хлорофилла вдоль неустойчивых многообразий гипер.
% Точек в окрест. вихря. Приведенный рис. Совершенно это не показывает.
% Нужно либо убирать этот параграф, который вызовет больше вопросов у рецензентов и
%читателей, чем даст ответов. Либо строить другой рис. не с FTLS с сайта АВИЗО а
%c собственной картой FTLE с наложенным хлорофиллом).
%

Mesoscale anticyclonic eddies with high primary production in the eastern subarctic Pacific and
the eastern BS region are able to influence the zooplankton which could, in turn, support higher
trophic levels and create favorable fishing grounds. Distribution, migration paths and growth
rate of salmon during its sea period of life are defined by oceanographic conditions at feeding
grounds. Mesoscale activity (the scale and strength of eddies, intensity of mesoscale water
transport, etc.) is one of the main factors which determines the dimension and spatial structure
of salmon feeding grounds \citep[see, e.g.,][]{Sobolevsky1994}. Formation and dynamics of the
salmon feeding base are influenced by the mesoscale activity in the region.

Eastern subarctic Pacific and the eastern BS are the main feeding areas for salmon in the
northern North Pacific \citep[see, e.g.,][]{Myers2007, Sato2009}. Figures~\ref{fig7}b--d
demonstrate year-to-year changes in the WSC in November\mdash March (the 5-years running mean)
in the northern North Pacific and annual catch of chum salmon (b, d) and coho salmon (c).
The total catch of chum salmon in the western Alaska area was comparatively low from about
1950 to 1970. Catches increased dramatically in 1975--1995 but declined in the late 1990s
and slightly increased in 2005--2009 (Figs.~\ref{fig7}b and d). The total catches of coho
salmon in the central, southwestern and eastern Alaska region and of chum salmon in the
western BS were relatively high ($30\text{--}50$ thousands ton and $2\text{--}5$ thousands
ton) from 1980 to mid-1990s but significantly decreased to about $20$ thousands ton and
$0.5\text{--}1$ thousands ton in 2000--2010 (Fig.~\ref{fig7}c). Strong correlations between
the changes in the chum and coho salmon catches in the eastern subarctic Pacific and the BS
and the WSC in the northern North Pacific in winter with the coefficient $r=0.64\text{--}0.75$
could be related to changes in the mesoscale eddy activity in the study area.

\section{Discussion}

The altimetry-based daily computed Lagrangian maps allow tracking of the origin and transport
pathways of the AS waters in the northern North Pacific and the BS and facilitate visualization
of mesoscale eddies in the study area. An intensification of the AS flow was observed in
November\mdash March when the Aleutian Low developed in the northern North Pacific, and strong
positive WSC appeared in the subarctic North Pacific and the BS. In spring\mdash fall, the
southward meanders of the AS and the flow of the AS water and open-ocean subarctic water
through the Aleutian passes into the BS were caused by mesoscale eddy activity.

Periods of increased eddy activity along margins of the eastern subarctic Pacific
may be related to anomalous downwelling wind conditions along the continental margin
\citep[see, e.g.,][]{Combes2007, Henson2008}. Increased poleward (downwelling favourable)
wind stress, increased WSC along the eastern boundary and subsequent increase in the Alaska
Current transport and intensification of the cyclonic gyre generate more mesoscale eddies
\citep{Okkonen_2001, Melsom2003, Ladd2007}. Wind stress curl is expected to be the most important
forcing factor for the large-scale variability of circulation in the eastern subarctic Pacific
\citep{Cummins1988}. \citet{Combes2009} concluded that on interannual and longer time scales,
the offshore transport of passive particles in the Alaskan Stream does not correlate neither
with a large-scale atmospheric forcing, nor with local winds. In contrast, in the Alaska
Current region, stronger offshore transport of passive particles  coincides with periods
of stronger downwelling which triggers development of stronger eddies.

Our results show that strength of the AS anticyclonic eddies (SSH in the eddy's center
and velocities at the boundaries) to the south off the Alaskan Peninsula and the eastern
Aleutian Islands are determined by the WSC in the northern North Pacific in November\mdash
March. One may assume that that spin-up of the cyclonic gyre in the subarctic Pacific,
forced by the WSC in November\mdash March, causes enhanced eddy activity along the continental
slope of the Alaskan Peninsula and the Aleutian archipelago. The annual modulation in the
cyclonic gyre's intensity in these areas should be interpreted as a barotropic response to
the seasonal WSC forcing. The magnitude of this response is quantifiable by a time-dependent
Sverdrup balance \citep{Bond1994, Ishi_2005}. Reinforcement and strengthening of the AS eddies
occur in the \NN{53}{55}, \WW{156}{158} area. The correlation between the SSH in the eddies,
the velocities at their boundaries and the WSC in the northern North Pacific in winter is
significant during two years while the eddies have been advected westward. An intensification
of the anticyclonic eddies activity is accompanied by an increase of the northward advection
of the open-ocean water to the eddy's boundaries and, thereby, an increase in the density
difference between anticyclonic eddies and ambient waters.

The significant correlation between the surface velocities at the outer shelf margin of the
eastern BS in summer and fall and the WSC in the North Pacific in winter (Fig.~\ref{fig3S}b) is related
probably to the AS and open-ocean water supply to the BS. The increased inflow through the
Aleutian passes may enhance eddy variability along the Bering Slope Current by increasing a
baroclinic instability \citep{Mizobata2008}. The supply of low salinity Alaska coastal waters
and relatively high salinity open-ocean waters creates zones with significant horizontal
salinity and density gradients and, thereby, could enhance the mesoscale dynamics
in the eastern BS.

An increase of the WSC in winter in the North Pacific activates anticyclone eddies in
the central part of the deep BS in summer and fall (Fig.~\ref{fig3S}a) and (with a 1-year lag)
the anticyclone eddies located in the Aleutian North Slope Current area (Fig.~\ref{fig3S}c). Our
results support the conclusion of \citet{Ladd_Stabeno_Ohern_2012} who indicated that the
anticyclonic eddy activity along the eastern shelf-break of the BS (the Pribilof eddy)
during the spring months is negatively correlated with the North Pacific Index, a measure
of the strength of the Aleutian Low in November\mdash March. \citet{Ladd_Stabeno_Ohern_2012}
assumed that a spin-up of the subpolar gyre in the northern North Pacific leads to increased
eddy activity in the BS, possibly due to a local effect of the stronger Bering Slope Current
or due to increased flow through the Aleutian passes.

Biological production in the deep basin of the BS is iron limited while the surface waters
at the shelf are iron replete and nitrate limited \citep{Aguilar2007}. High surface \chla
concentrations along the shelf break in the BS appears to be associated with an eddy-induced
mixing between shelf and deep basin waters
\citep[see, e.g.,][]{Okkonen2004, Mizobata2008, Ladd_Stabeno_Ohern_2012}. The mixing, induced
by anticyclonic eddies between the low salinity coastal water and high salinity deep basin water,
probably creates favourable conditions with a nitrate availability and a shallow pycnocline for
the phytoplankton growth significantly increasing concentrations of \chla in the upper surface
layer in the eastern BS in summer (Figs.~\ref{fig3S}a,~b and~d). The canyons, located at the shelf break
(the Bering, Pribilof, Zhemchuk and Navarin ones), are considered to be preferred
sites of cross-shelf exchange \citep{Clement_Kinney2009}.

An increase of the WSC in the northern North Pacific in winter impacts the mesoscale dynamics
in the eastern subarctic Pacific and the eastern BS (Figs.~\ref{fig4} and~\ref{fig3S}) and, thereby,
a \chla concentration in surface waters (Fig.~\ref{fig6}). Enhanced phytoplankton productivity
in anticyclonic eddies may transfer up the food chain and results in increased biomass of higher
trophic levels (zooplankton and fish). Advection paths of eggs and larvae can influence fish growth,
survival and recruitment, either propelling them towards areas that support high growth and survival,
or diverting them away from suitable habitat. For many fish species that spawn along the east
subarctic continental slope, eggs and larvae benefit from slope to continental-shelf transport,
where larvae encounter favorable feeding and growth conditions prior to the onset of winter
\citep{Bailey2008, Atwood2010}. General biological efficiency, defined by success of reproduction
of organisms at the lowest trophic levels, has priority value for formation of steady
salmon feeding conditions.

In the western part of the BS the highest biomass of chum salmon has been observed at
periphery of anticyclonic eddies where zooplankton is accumulated in the upper 100-meter layer
\citep{Sobolevsky1994}. \citet{Moss2013} demonstrated that salmon, caught along the anticyclonic
eddy's periphery in the eastern subarctic Pacific (the Sitka eddy), displayed the highest levels
of insulin-like growth factor which is an index of the short-term growth rate for salmon.
Zooplankton and phytoplankton densities are also greatest at the eddy's periphery. The location,
timing and strength of the Sitka anticyclonic eddy, combined with juvenile salmon outmigration
timing, could positively affect the growth by increased foraging opportunities. Years with
enhanced production at the eddy's periphery and reduced inter- and intra-specific competition,
resulting in increased survival for certain stocks \citep{Moss2013}, are those when the three
primary eddy features in the eastern subarctic Pacific (the Haida, Sitka and Yakutat eddies)
have been located close to the shore during early summer months when juvenile salmon are
migrating to the north. The observed correlations between the WSC and salmon catches for
the eastern subarctic Pacific and the eastern BS area (Figs.~\ref{fig7}b--d) may be explained
by an intensification (slow down) of the mesoscale dynamics and cross shelf exchange by macro-
and micro- nutrients, stimulating phytoplankton and zooplankton growth forced by increased
(decreased) WSC in the northern North Pacific in winter.

\section{Conclusions}

In this paper we proposed the forcing pattern that contributes to the interannual variability
of the mesoscale dynamics and the \chla concentration in the surface waters in the Alaskan Stream
area and the eastern Bering Sea. We conclude that the strength of the anticyclonic eddies along
the deep basin slopes of the northern subarctic Pacific and the eastern Bering Sea is determined
by the wind stress curl in the northern North Pacific in November\mdash March. Strong correlations
have been found between the concentrations of \chla at the shelf\mdash deep-sea boundaries of the
Bering Sea and the northern Pacific subarctic in August\mdash September and the wind stress curl
in the northern North Pacific in November\mdash March. Our results indicate that the mesoscale
dynamics in the eastern subarctic Pacific and the eastern Bering Sea areas may determine not only
lower-trophic-level organism (autotrophic phytoplankton) biomass but also salmon abundance and catch.

\section*{Acknowledgements}
The work was supported by the Russian Science
Foundation (project no.~16--17--10025) and its methodological part was supported by the FEBRAS
Program (No. AAAA-A17-117030110034-7). The altimeter products were distributed by AVISO with
support from CNES.

\pagebreak
\appendix\section{Supplementary materials}
\subsection{Origin Lagrangian maps in the study area}
\begin{figure*}[!hb]
\begin{center}
\includegraphics[width=0.49\textwidth,clip]{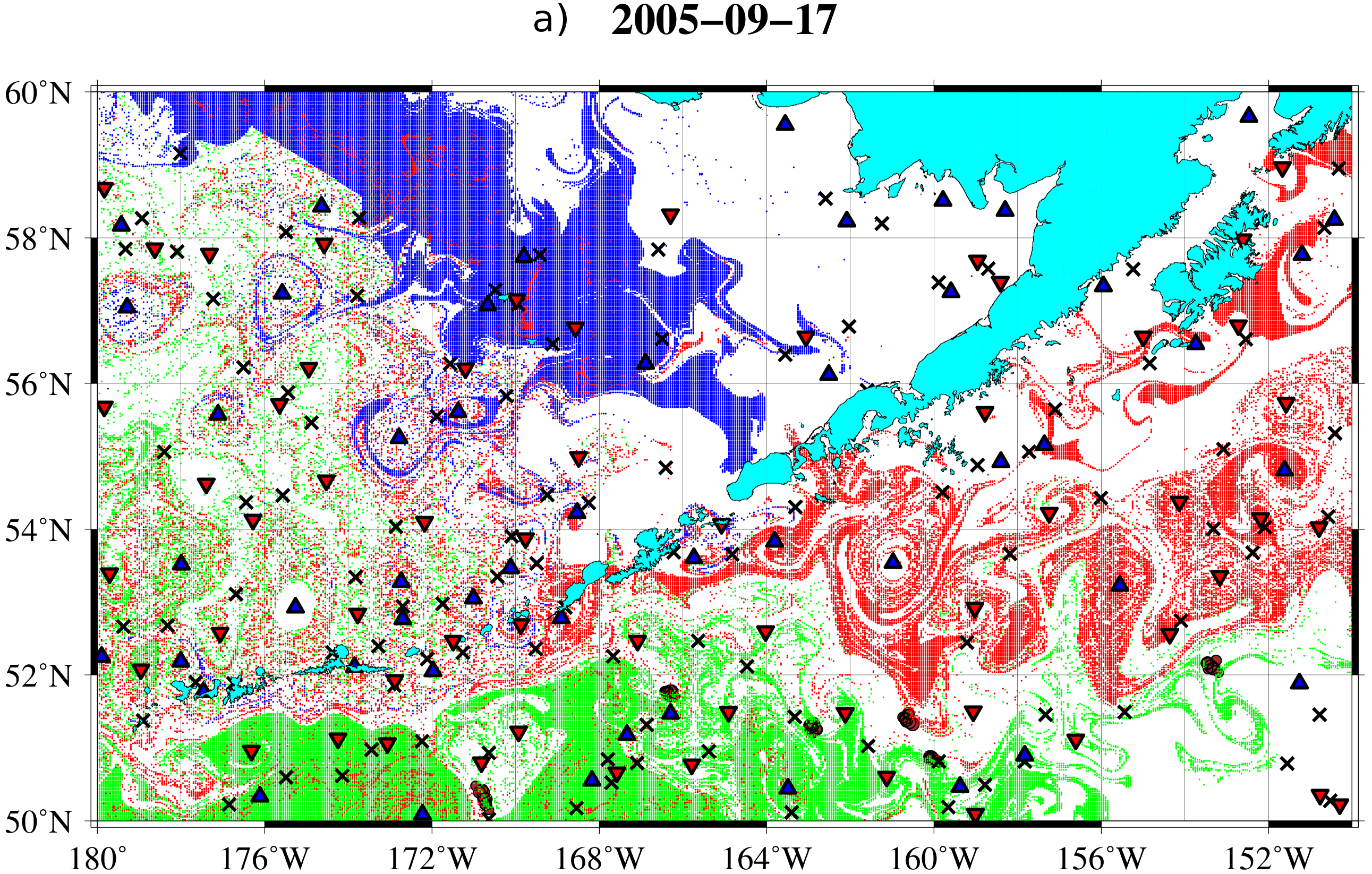}
\includegraphics[width=0.49\textwidth,clip]{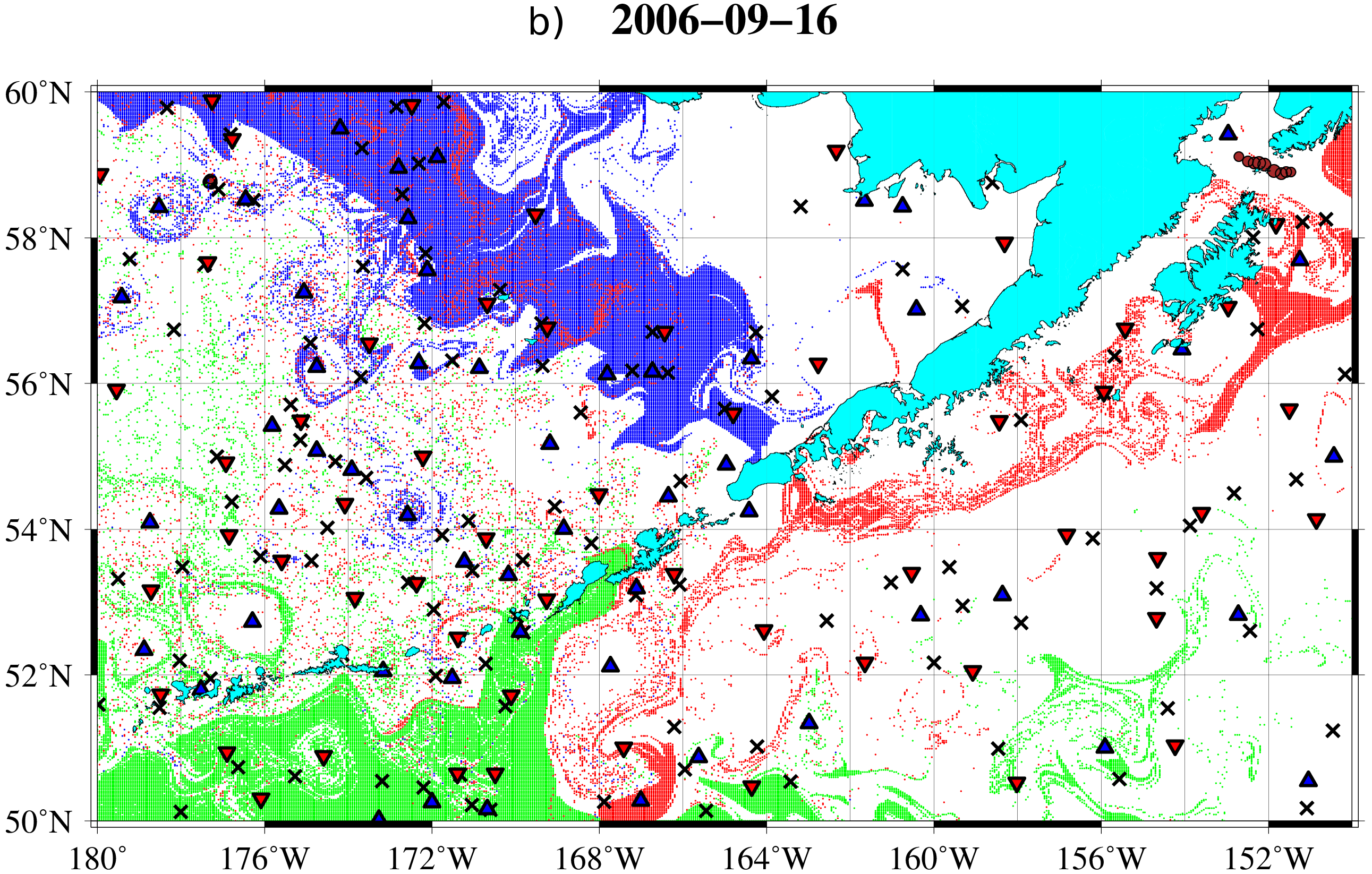}\\
\includegraphics[width=0.41\textwidth,clip]{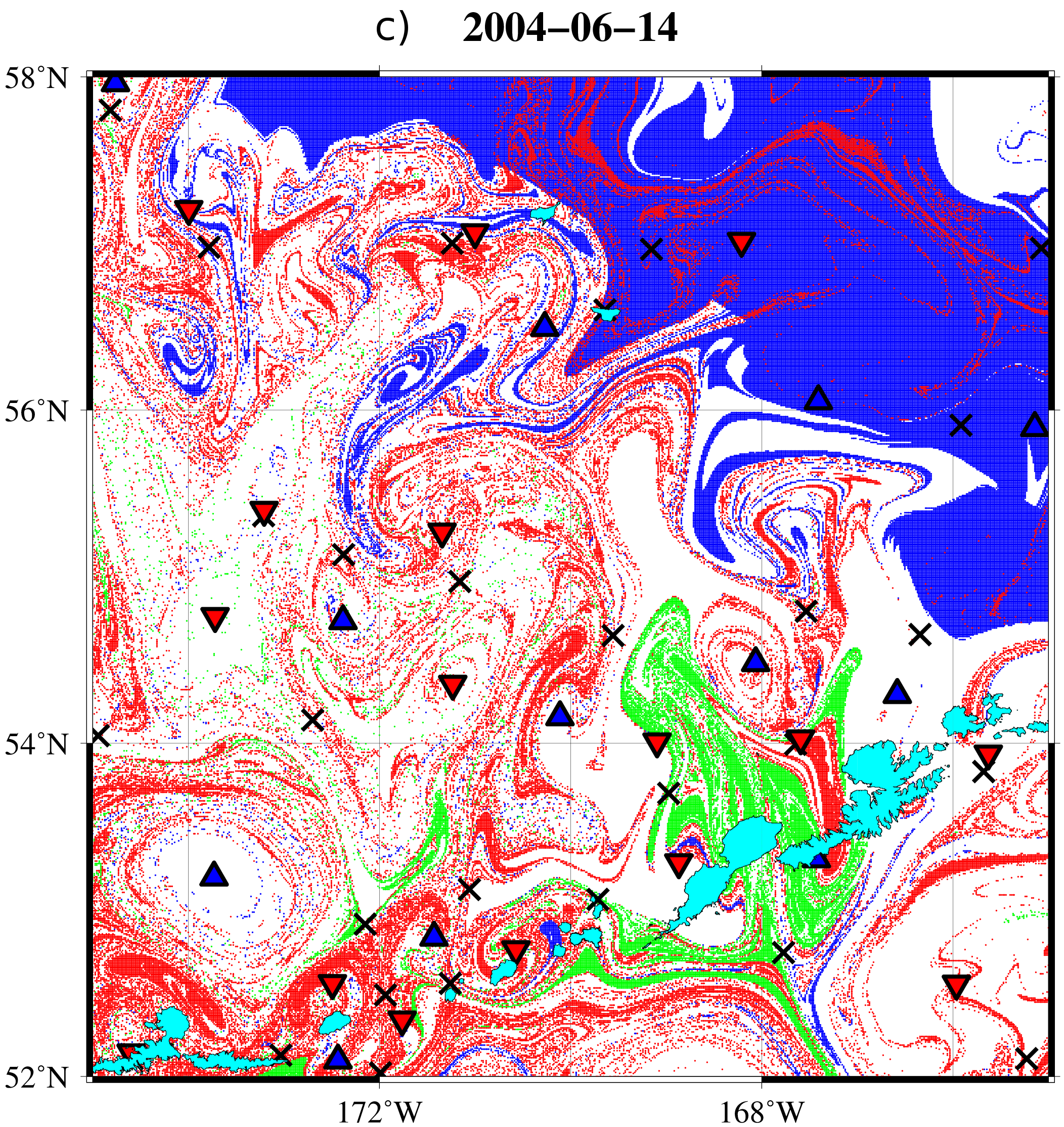}
\includegraphics[width=0.41\textwidth,clip]{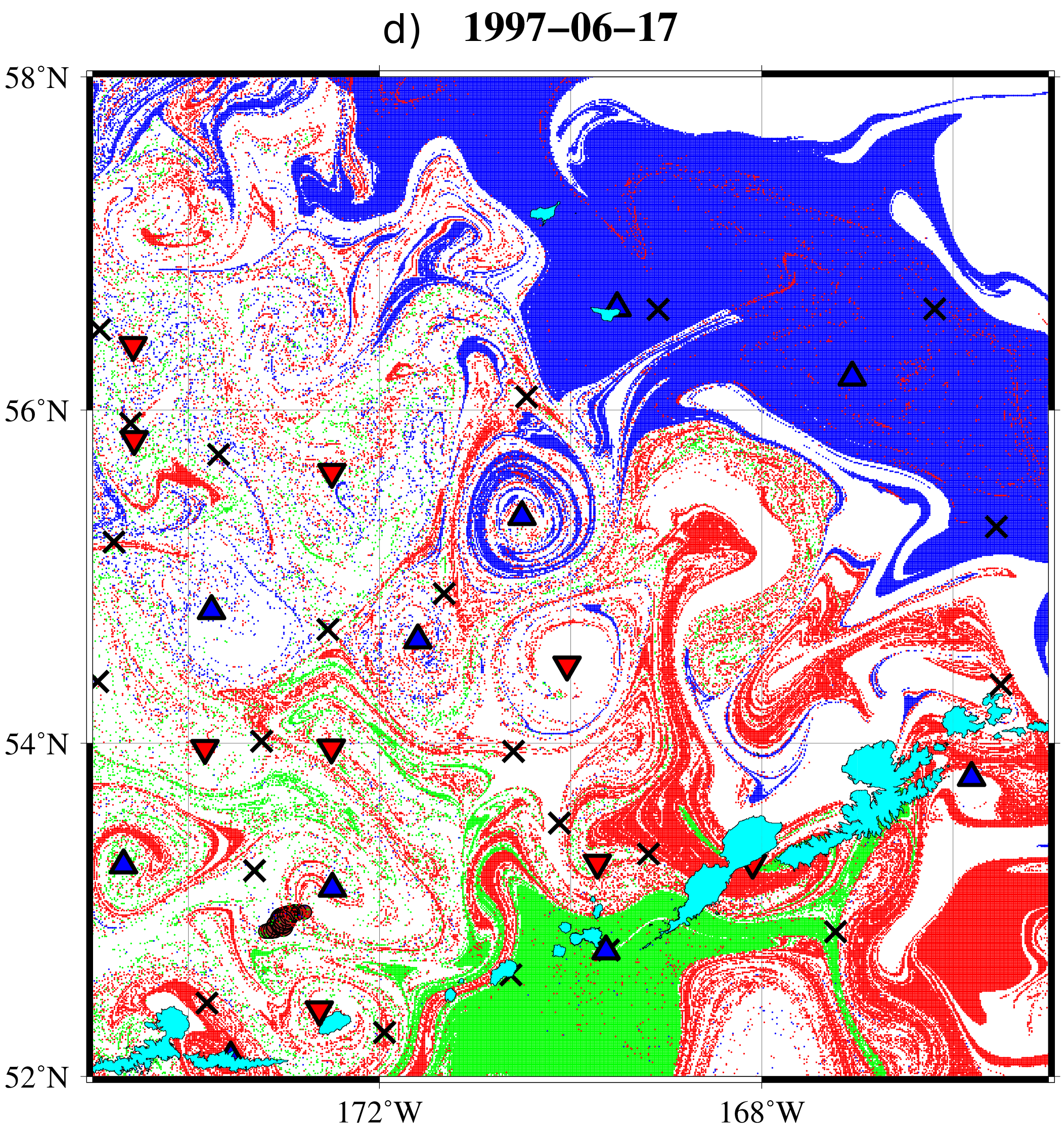}\\
\end{center}
\caption{The Lagrangian maps show transport pathways, origin, history and fate of Alaskan Stream
(AS) (red), open-ocean (green) and Bering Sea (BS) (blue) waters in a) September 2005, b) September
2006, c) June 2004 (the center of the Pribiloff mesoscale anticyclone is at the point \N{54.5}, \W{168})
and d) June 1997 (the center of the Pribiloff mesoscale anticyclone is at the point \N{55.5}, \W{171}).
The penetration of the BS shelf waters into the deep basin of the eastern BS is demonstrated by the
blue color. Elliptic and hyperbolic stagnation points with
zero velocity are indicated by triangles and crosses, respectively.}
\label{fig1S}
\end{figure*}

We are interested in three different water masses and their transport pathways. To track the Alaskan Stream
(AS) waters, the section along the meridian $x_0=\W{145}$ from $y_0=\N{58}$ to $y_0=\N{60}$ is fixed.
The particles, which crossed that section in the past, are colored in red on the origin Lagrangian maps.
The open-ocean particles, which crossed the section $x_0=\EW{160.0}{164.0}$, $y_0=\N{50.0}$ in the
past, are colored in green. The eastern Bering Sea (BS) particles, which crossed the section from
\E{177.0}, \N{62.0} to \W{164}, \N{55.0} in the past, are colored in blue (see the yellow line in
Fig.~\ref{fig2}a). We removed from consideration all the particles entered into any AVISO grid
cell with two or more corners touching the land in order to avoid artifacts due to the inaccuracy
of the altimetry-based velocity field near the coast. The corresponding colored Lagrangian maps
in Fig.~\ref{fig1S} demonstrate clearly origin, history and fate of those water masses in the study area.

\pagebreak
\subsection{Vertical distributions of temperature, salinity and potential density inside and outside
of the Alaskan Stream anticyclone in 2005--2006}
\begin{figure*}[!hb]
\begin{center}
\includegraphics[width=0.8\textwidth,clip]{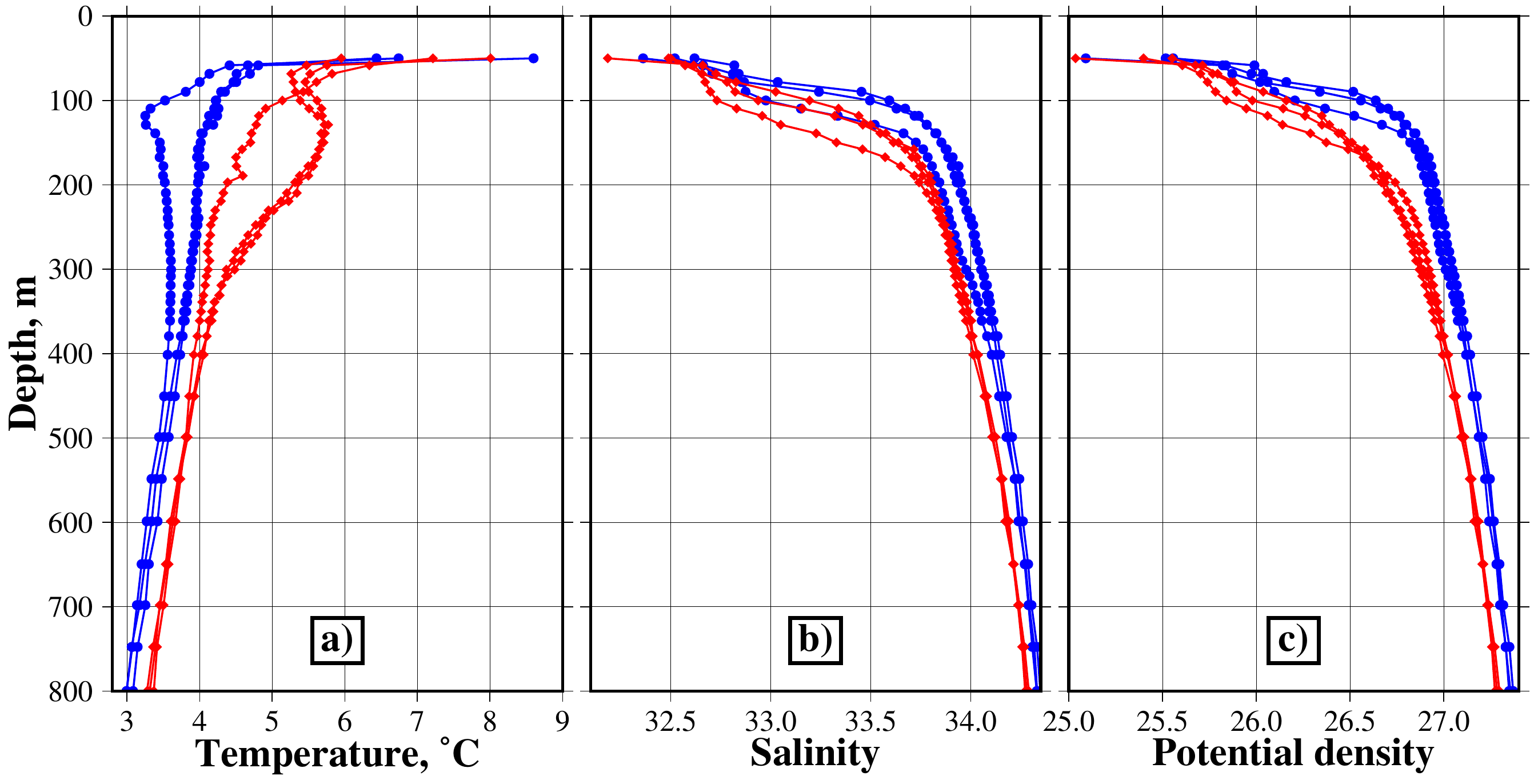}
\end{center}
\caption{The vertical distributions of temperature, salinity and relative density
in September 2005 and 2006
inside (red color) and outside (blue color) the Alaskan Stream  anticyclone 2005--2006 centered at around
\N{54}, \W{161} in Fig.~\ref{fig1S}a on
17 September 2005 and at \N{51}, \W{167} in Fig.~\ref{fig1S}b on 16 September 2006.
The data were taken from the Argo buoys nos. 4900342, 4900397, 4900646 and 4900705.
Similar to ASAC~2003--2004 (see Figs.~\ref{fig2}d--f), the ASAC~2005--2006 core was composed
of relatively low salinity (33.7--33.9) and low density (26.7--26.9) waters. The temperature of waters
inside of the anticyclone was 1--2~$^\circ$C higher than outside it.}
\label{fig2S}
\end{figure*}

\pagebreak
\subsection{Annual changes of the wind stress curl in the northern North Pacific and the meridional and zonal velocities in the eastern Bering Sea}
\begin{figure*}[!hb]
\begin{center}
\includegraphics[width=0.8\textwidth,clip]{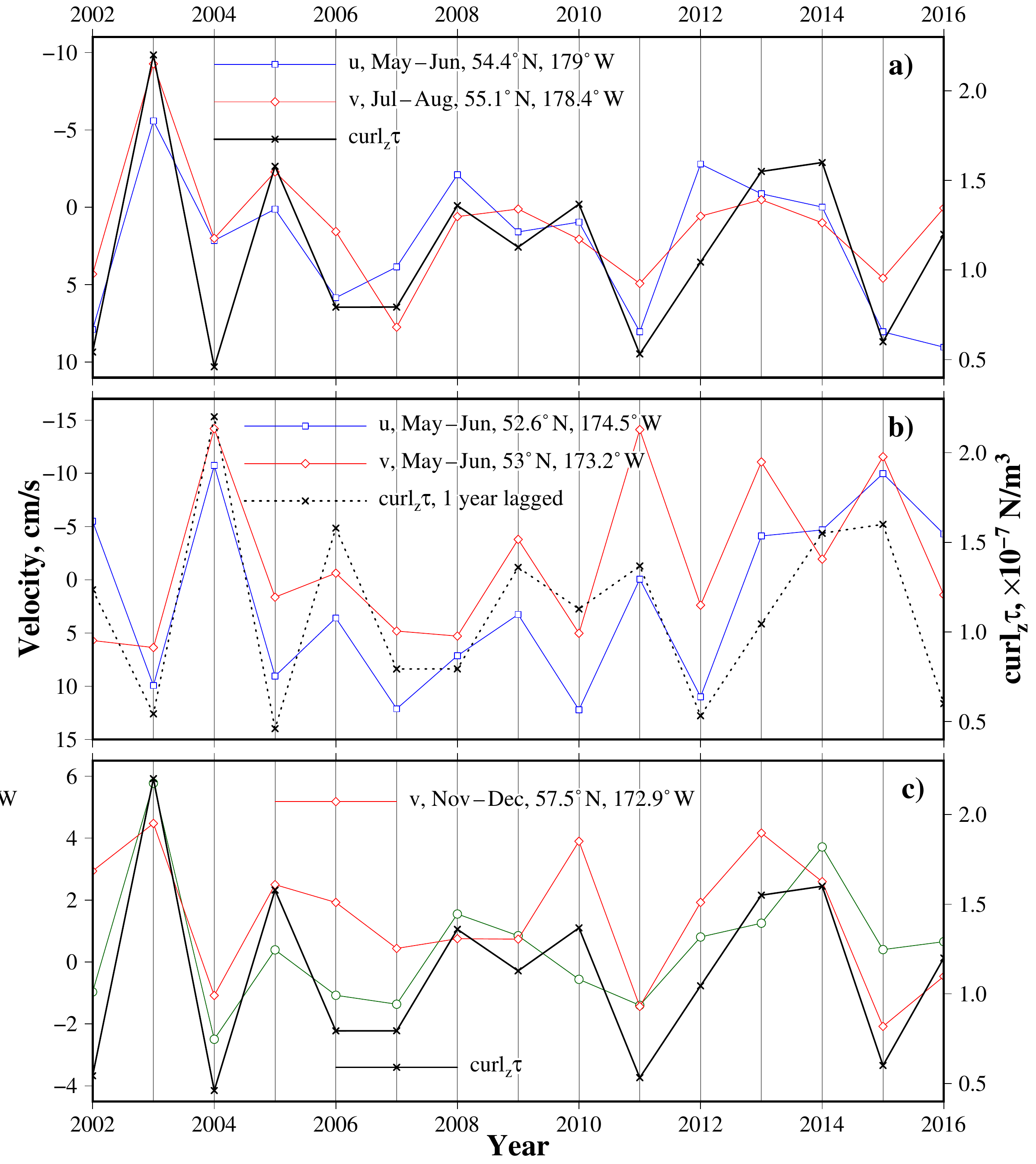}
\end{center}
\caption{a), b) and c) The year-to-year changes of the wind stress curl
(November\mdash March) in the northern North Pacific
and the meridional and zonal velocities in the eastern Bering Sea.}
\label{fig3S}
\end{figure*}

The changes in the Aleutian Low activity and the wind stress curl in the northern North Pacific
in winter determine year-to-year changes in velocities in some areas of the eastern Bering Sea
(see Fig.~\ref{fig4}). An increase (decrease) of the WSC in the
North Pacific in November\mdash March is accompanied by increased (decreased) velocities at
the boundaries of the  anticyclonic eddies in the central part of the deep Bering Sea in summer
and fall (Fig.~\ref{fig3S}a). An intensification of
the Aleutian Low and a large positive wind stress curl result in increasing of the northward
flow on the Bering Sea outer shelf in the areas located close to the Pribiloff, Zhemchug and
Navarin canyons (Fig.~\ref{fig3S}b). An increase (decrease) of the wind stress curl in the northern
North Pacific in November\mdash March with a 1-year lag is accompanied by increased
(decreased) velocities at the boundaries of the anticyclonic eddies located in the area of
Aleutian North Slope Current in summer and fall (Fig.~\ref{fig3S}c).

\pagebreak
\subsection{Distribution of \chla concentration and salinity in the eastern Bering Sea}
\begin{figure*}[!hb]
\begin{center}
\includegraphics[width=0.89\textwidth,clip]{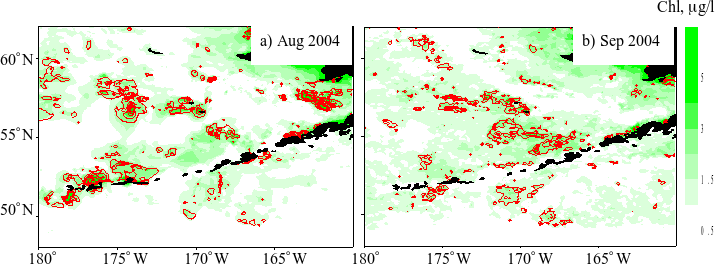}\\[5mm]
\includegraphics[width=0.4\textwidth,clip]{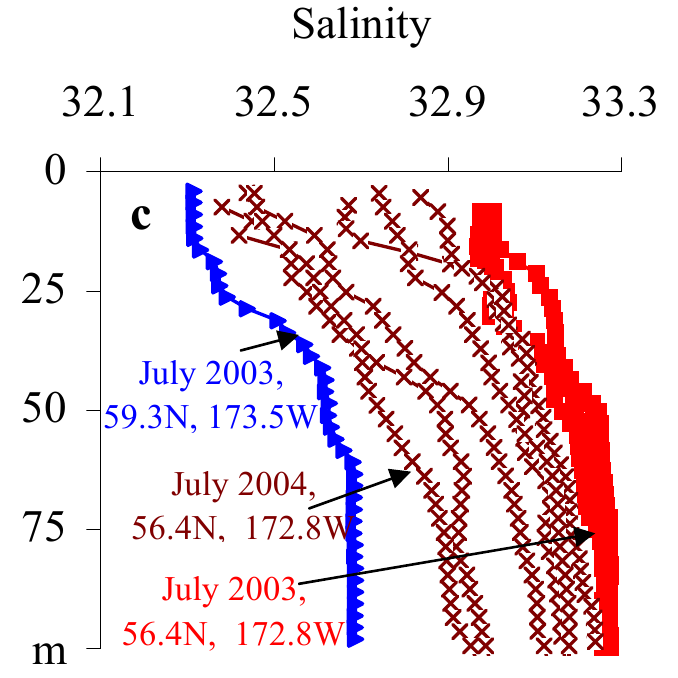}%
\includegraphics[width=0.4\textwidth,clip]{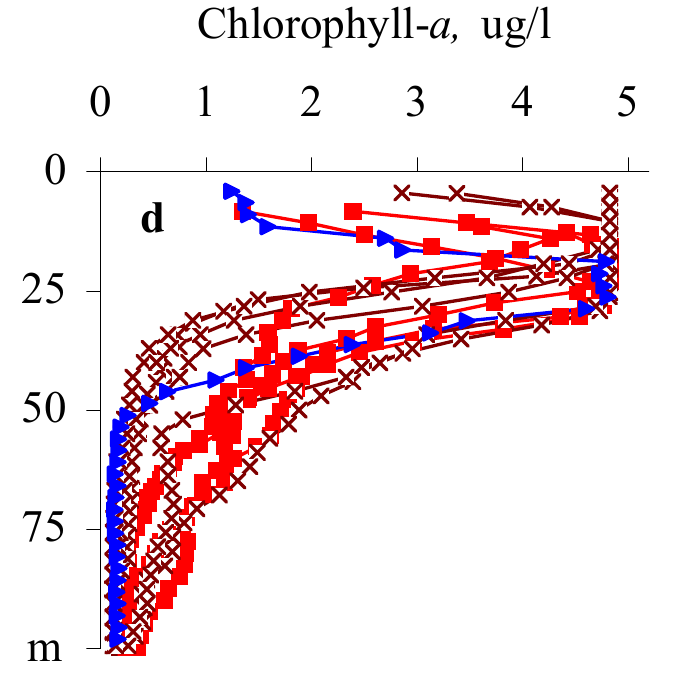}\\
\includegraphics[width=0.4\textwidth,clip]{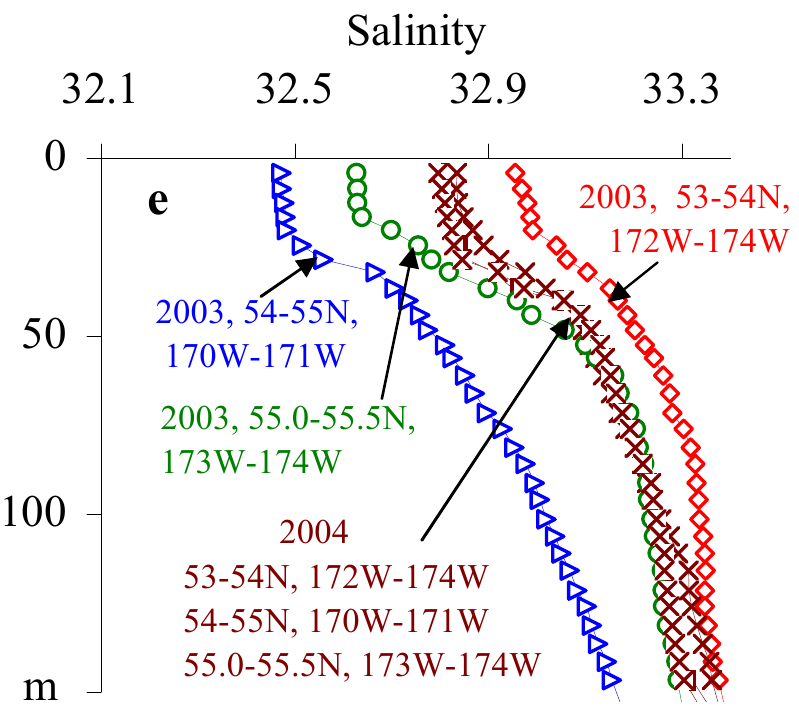}
\end{center}
\caption{a) and b) Distribution of the \chla concentration in August and September 2004 shown
by the green color and the difference in \chla concentration between 2004 and 2003 shown by the
red color lines; 1--5~$\mu$g/l with the interval equal to 1~$\mu$g/l. c) and d) The vertical distributions of salinity
and \chla in July 2003 and July 2004 in the eastern BS and e) the vertical distribution of salinity
in the eastern Bering Sea in July\mdash September 2003 and 2004 (the data from  the Argo buoys
nos.~4900142, 4900145, 4900165, 4900167 and 4900168).}
\label{fig4S}
\end{figure*}

\pagebreak
\subsection{The impact of anticyclonic eddies on the \chla distribution in the study area}
The impact of anticyclonic eddies on the \chla distribution (the MODIS data) in the AS area can be demonstrated
by using a Lagrangian indicator $L=\int\limits_{0}^{T} \sqrt{u^2+v^2}dt$ which
is a measure of a distance passed by advected  particles. A studied area has
been seeded with a large number of virtual particles whose trajectories
have been computed backward in time in the AVISO velocity field for a month from the date
indicated on the corresponding maps.
The $L$ maps visualize not only the very vortex structures but also a history of water masses to
be involved in the vortex motion in the past.
\begin{figure*}[!hb]
\begin{center}
\includegraphics[width=0.5\textwidth,clip]{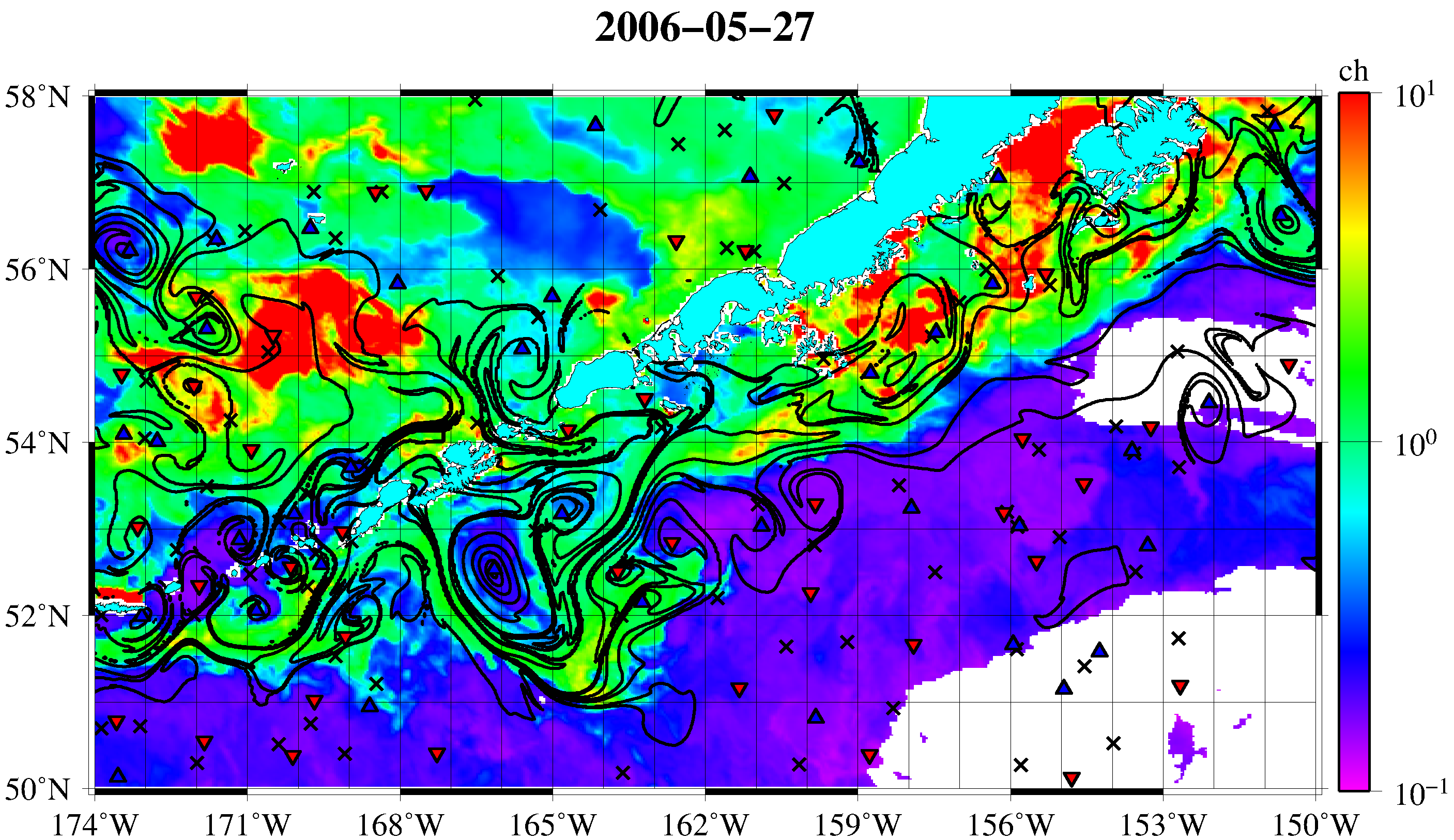}\\
\includegraphics[width=0.5\textwidth,clip]{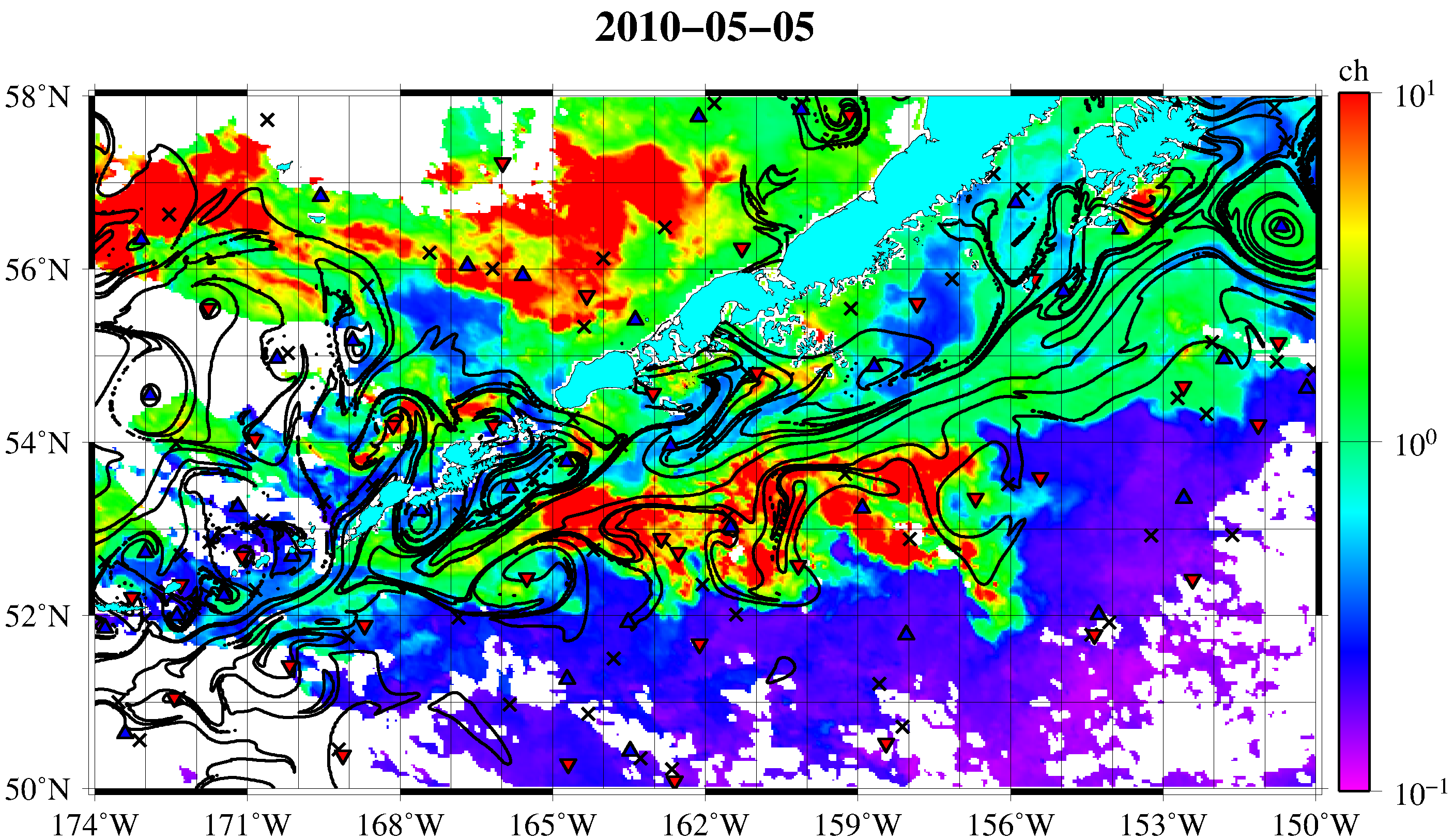}\\
\includegraphics[width=0.5\textwidth,clip]{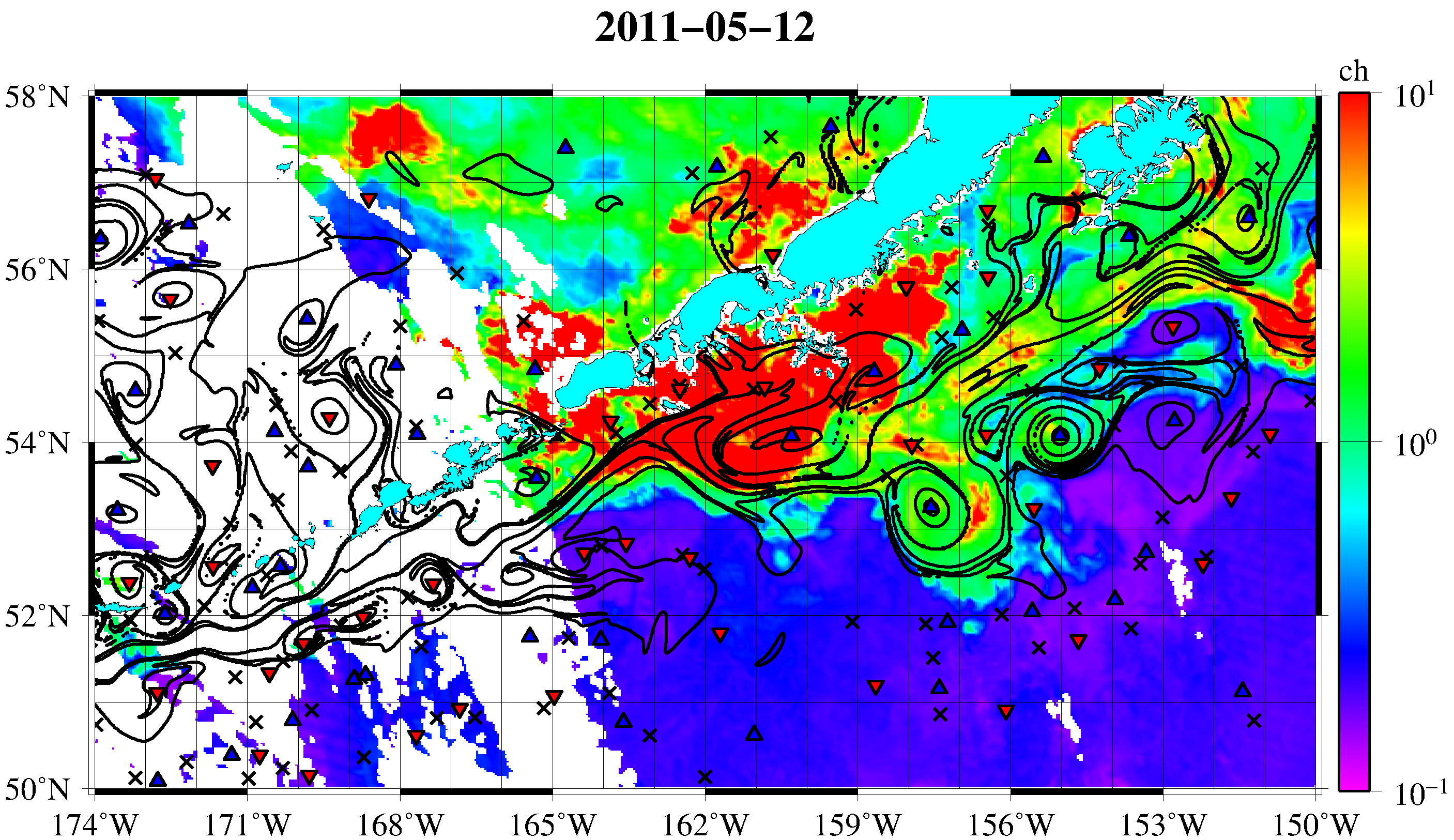}\\
\end{center}
\caption{The black isolines of the Lagrangian indicator $L$ with the step of 200 geographic minutes
imposed on the \chla distribution in the AS area (May 2006, May 2010 and May 2011). They
enclose stable mesoscale eddies, such as ones with the elliptic points at \N{52.5}, \W{165}
and at \N{53}, \W{164}. The dominant feature in the \chla distribution in the surface
layer is a contrast between coastal and offshore waters. The coastal waters are productive with
high values of \chla (${>}6$~$\mu$g/l), and the off waters are oligotrophic with low
\chla values (${<}1$~$\mu$g/l). The filaments with high \chla concentration are
wrapped around persistent mesoscale eddies.}
\label{fig5S}
\end{figure*}

\clearpage

\bibliographystyle{model2-names}
\bibliography{elsarticle-template-2-harv}{}

\end{document}